\documentclass [11pt,a4paper]{article}
\usepackage{jcappub}

\usepackage{amsmath,amsfonts,bbm}
\usepackage{graphicx}
\usepackage{url}
\usepackage{here}

\newcommand{\be}{\begin{equation}}
\newcommand{\ee}{\end{equation}}
\newcommand{\ba}{\begin{eqnarray}}
\newcommand{\ea}{\end{eqnarray}}
\def\bs{\begin{subequations}}
\def\es{\end{subequations}}
\def\a{\alpha}
\def\b{\beta}
\def\de{\delta}

\def\la{\lambda}
\def\k{\kappa}
\def\e{\epsilon}
\def\ve{\varepsilon}
\def\Om{\Omega}
\def\om{\omega}
\def\G{\Gamma}

\def\s{\sigma}

\def\cA{\mathcal{A}}
\def\cB{\mathcal{B}}

\def\cE{\mathcal{E}}

\def\cP{\mathcal{P}}

\def\cR{\mathcal{R}}

\def\cU{\mathcal{U}}
\def\cV{\mathcal{V}}

\def\dh{d_\textsc{h}}

\def\p{\partial}

\newcommand{\Eq}[1]{(\ref{#1})}
\def\com{\color{magenta}}
\def\cob{\color{blue}}

\newcommand{\oarX}[1]{\href{http://arxiv.org/abs/#1}{{\ttfamily\com #1}}}
\newcommand{\arX}[1]{\href{http://arxiv.org/abs/#1}{{\ttfamily\com arXiv:#1}}}
\newcommand{\doij}[5]{\href{http://dx.doi.org/#1}{\cob {\it #2} {\bf #3} (#5) #4}}
\newcommand{\doin}[6]{\href{http://dx.doi.org/#1}{\cob {\it #2} {\bf #3 #4} (#6) #5}}
\newcommand{\doinn}[5]{\href{http://dx.doi.org/#1}{\cob {\it #2} {\bf #3} (#5) #4}}

\newcommand{\tia}[1]{\textit{#1},}
\newcommand{\boxd}[1]{\boxed{\phantom{\Biggl(}#1\phantom{\Biggl)}}}
\def\Pl{{\rm Pl}}
\def\lp{\ell_\Pl}
\def\mpl{m_\Pl}
\def\tp{t_\Pl}
\def\rme{e}
\def\rmd{d}

\begin{document}

\title{Cosmic microwave background and inflation in multi-fractional spacetimes} 

\author[a]{Gianluca Calcagni,}
\affiliation[a]{Instituto de Estructura de la Materia, CSIC, Serrano 121, 28006 Madrid, Spain}
\emailAdd{calcagni@iem.cfmac.csic.es}

\author[b,c]{Sachiko Kuroyanagi,}
\affiliation[b]{Department of Physics, Nagoya University, Chikusa, Nagoya 464-8602, Japan}
\affiliation[c]{Institute for Advanced Research, Nagoya University, Chikusa, Nagoya 464-8602, Japan}
\emailAdd{skuro@nagoya-u.jp}

\author[d]{Shinji Tsujikawa} 
\affiliation[d]{Department of Physics, Faculty of Science, Tokyo University of Science, 
1-3, Kagurazaka, Shinjuku-ku, Tokyo 162-8601, Japan}
\emailAdd{shinji@rs.kagu.tus.ac.jp}

\abstract{We use FIRAS and \textsc{Planck} 2015 data to place observational bounds on inflationary scenarios in multi-fractional spacetimes with $q$-derivatives. While a power-law expansion in the geometric time coordinate is subject to the usual constraints from the tensor-to-scalar ratio, model-independent best fits of the black-body and scalar spectra yield upper limits on the free parameters of the multi-fractal measure of the theory. When the measure describing the fractal spacetime geometry is non-oscillating, information on the CMB black-body spectrum places constraints on the theory independent from but weaker than those obtained from the Standard Model, astrophysical gravitational waves and gamma-ray bursts (GRBs). When log oscillations are included and the measure describes a discrete fractal spacetime at microscopic scales, we obtain the first observational constraints on the amplitudes of such oscillations and find, in general, strong constraints on the multi-scale geometry and on the dimension of space. These results complete the scan and reduction of the parameter space of the theory. Black-body bounds are obtained also for the theory with weighted derivatives.}




\maketitle

\section{Introduction}

The satellite \textsc{Planck} has crowned the age of precision cosmology inaugurated by \textsc{cobe} \cite{Fix96} for the observation of the Cosmic Microwave Background (CMB). The \textsc{Planck} 2015 determination of the cosmological parameters has reached the level of several digits \cite{P1513,P1520} and, in cooperation with the bounds from other experiments such as \textsc{Bicep2} \cite{Bic15}, it has placed severe restrictions on theoretical models of the early universe, to the point where quadratic inflation is excluded at the 95\% confidence level. The data allow us to constrain several different scenarios, for instance from string theory, supergravity and quantum gravity \cite{MRV,infcon1,infcon2}. In this paper, we will examine a theory of multi-fractal spacetimes which predicts characteristic features in the power spectrum. Multi-scale spacetimes are continuum geometries whose properties vary with the scale of experiments \cite{fra1,fra4,frc1,frc2,frc12,frc13,trtls,qGW} (see \cite{frc11} for a somewhat outdated overview). In some special cases, such as the model with $q$-derivatives we will consider here, these geometries can be regarded as approximations of \emph{multi-fractals}. Multi-scale spacetimes are the generalization of fractal sets to Lorentzian geometries. A fractal is a nowhere-differentiable set of points whose dimensionality can be non-integer (and varying with the scale, in the case of multi-fractals). To recall the example which gave birth to fractal geometry \cite{Man67}, the Western coast of Great Britain is so irregular that it has infinite length if measured with infinitesimal standard linear rods. However, its geometry turns out to be well defined when measured with ``anomalous'' rods. Its dimension is between 1 and 2, so that it is ``more'' than a smooth curve but it does not fill the plane completely. Modulo several important distinctions, the notion of fractal can be extended to space and time \cite{trtls}.

The motivation to look beyond standard manifolds is twofold. First, anomalous geometries and anomalous transport phenomena are the rule rather than the exception in Nature. Complex dynamics can emerge in systems which, at the microscopic level, are described by non-chaotic physics on smooth geometries. In such effective regimes, it may be necessary to change the description of the system by employing different mathematical tools. Consider, for instance, an attempt made in cosmology in relation with dark energy, by explaining it not as a matter degree of freedom (quintessence) but as the manifestation of a geometry beyond general relativity. Modified gravity and $f(R)$ models do not avoid problems of naturalness and fine tuning in general. One may therefore be curious about exploring alternative theories and check whether observations and conceptual issues can be reconciled simultaneously. Multi-scale spacetimes, where anomalous scalings are implemented by default and entail a much more radical depart from general relativity than modified-gravity models, have several interesting features which can have applications to cosmology but they have been studied only from a theoretical point of view \cite{fra2,frc11}. CMB data offer a unique opportunity to check up to what scales effects of fractal geometry are negligible and, therefore, to constrain the scale hierarchy of the model.

Second, a change of dimensionality of spacetime has been recognized as an almost universal feature of theories of quantum gravity \cite{tH93,Car09,fra1}, as several examples in quantum gravity show (we mention, among many others, causal dynamical triangulations, asymptotic safety, Ho\v{r}ava--Lifshitz gravity and non-commutative spacetimes). This dimensional flow is, in some instances, related to good ultraviolet properties (renormalizability or absence of infinities) and it is therefore important to understand such relation. To do so, it is very useful to import the techniques of fractal geometry and anomalous transport theory to the realm of particle physics and quantum gravity. Multi-scale spacetimes incorporate these techniques while at the same time avoiding the formidable technical challenges one must face when building a quantum field theory directly on disconnected fractal structures \cite{Sti77,Svo87,Ey89a,Ey89b}. Multi-scale models offer either an effective description of certain quantum-gravity theories in the limit of a continuum commutative manifold \cite{ACOS,fra7}\footnote{This as a connection with the ``change of rods'' of the first motivation \cite{fra7,trtls}.} or a stand-alone proposal to test against observations. In this paper, we will adopt the latter view.

Multi-scale geometries are characterized by logarithmic oscillations in the measure, which is inherited by a log-periodic dependence of the inflationary power spectrum from the wave-number. This feature was briefly noted in \cite{frc11} but we will present it here in greater detail. In comoving momentum space, log oscillations
\be\label{heurlog}
A\cos(\om\ln k)+B\sin(\om\ln k)
\ee
signal the transition from a continuous geometry at mesoscopic and large scales to a discrete geometry at microscopic scales. The frequency of oscillations $\om$ sets the ratio at which the geometry is self-similar \cite{frc2}. Cosmological observations can give an estimate of this frequency and, therefore, important information on the discrete-to-continuum transition in multi-fractional spacetimes. Another feature typical of multi-scale spectra is that the spectral index effectively changes with the scale. Thus, another question we would like to pose is whether this change, or its combination with log oscillations, can explain the fluctuations in power at large multipoles with respect to the $\Lambda$CDM model. These features are compatible with the base cosmological model within the error bars; nevertheless, it may be interesting to investigate them further.

Log-oscillating inflationary spectra, where corrections to the standard spectrum take the form \Eq{heurlog} in the comoving wave-number $k$, are not new in the literature. They can arise from different mechanisms: in generic trans-Planckian models where perturbation modes are created in a vacuum with a characteristic (or ``preferred'') scale \cite{MB1,BM1} (tested with WMAP \cite{MaRi1,MaRi2,EKP,HHSW} and first-year \textsc{Planck} data \cite{HHSW}), in ``ordinary'' (i.e., non-multi-scale) cyclic inflation \cite{BMSh}, by resonances in multi-field inflation \cite{Che11,BNV}, in axion monodromy inflation \cite{FMPWX} and in ``unwinding'' inflation \cite{DGKS}, the latter two models having been already constrained with \textsc{Planck} data, both with the first release \cite{MSW1,MSW2} and the second \cite{P1520}. Other types of oscillatory spectra, with linear arguments of the periodic functions (corrections of the form $\cos(\om k)+\sin(\om k)$), can be produced in effective field theory \cite{GSvS}, by generic sharp features in the inflaton potential \cite{CEL,HCMS} or in D-brane inflation in string theory, where their origin can be traced back to features in the warp factor of type-IIB Calabi--Yau flux compactifications \cite{BCHTX}. The shape of the fit function in all these cases differs from ours and the multi-fractional theory will require a brand new numerical analysis. 

It is important to stress that, although the idea of constraining the dimensionality of spacetimes via experiments is not new, there are no constraints on realistic multi-scale models apart from those found in \cite{frc12,frc13,qGW} very recently. In fact, previous studies\footnote{The reader can find an exhaustive list of references and of their results in section 3.4 of \cite{frc2}.} were based on a generic ``dimensional regularization'' scheme (not motivated by any theoretical framework) where the dimension of spacetime is $D=4-\e$ across all scales. In this case, the parameter $\e$ must be extremely small, since it must undergo a battery of tests ranging from atomic to cosmological scales. In particular, in \cite{CO} the data collected by the FIRAS instrument on \textsc{cobe} \cite{Fix96} were used to estimate the allowed deviation of Planck's law (the frequency distribution of the radiance of a black body) in $3-\e$ dimensions with respect to the three-dimensional distribution so precisely determined for the CMB radiation. The result is the upper bound \cite{CO}
\be\label{boue3}
|\e|<10^{-5}\,,
\ee
valid at all scales comprised between the decoupling era and today. Although much less severe than the $|\e|<10^{-11}$ bound from the Lamb shift in hydrogen atoms \cite{ScM,MuS}, this constraint is stringent enough to conclude that no-scale toy models of dimensional regularization do not leave an appreciable imprint in the CMB. On the other hand, \emph{multi}-scale models easily avoid these conditions, since bounds found in an experiment conducted at a given scale may not apply to the physics of such models at other scales. We will show this explicitly by working out CMB-related bounds and comparing them with those from particle physics \cite{frc12,frc13}, gravitational waves and GRBs \cite{qGW}. Therefore, while there is a precise mapping between dimensional-regularization toy models and rigorous fractional (no-scale) theories \cite{frc2}, bounds such as the one above must be reworked completely anew in the case of multi-scale spacetimes, in particular for their realization with a multi-fractional measure.

Our results will show that, when log oscillations are ignored, information on the CMB black-body and inflationary spectra is not sufficient to constrain multi-fractional models better than Standard Model experiments \cite{frc12} and observations of astrophysical gravitational waves and GRBs \cite{qGW}. In fact, the CMB energy scale is too low to force a useful lower bound on the characteristic energy scale at which fractal effects leave an imprint. Our lower bounds will be in fact around $E_*>1- 100\,{\rm GeV}$, much lower than the current center-of-mass energy at the LHC. Nevertheless, translating this result into bounds on the fundamental length and time scales of the geometry and comparing it with the particle-physics bounds of \cite{frc12}, we find that these independent bounds are similar. Our Monte Carlo simulations are less effective in placing constraints on the parameters of theory due to a degeneracy in the ways viable spectra can be obtained from the parameter space. This feature is all the more important in the presence of log oscillations\footnote{To deal with oscillating likelihood profiles (as when we keep the frequency $\om$ free in a preliminary analysis), we include the \textsc{CosmoChord} plug-in \cite{cocor} in the \textsc{CosmoMC} code to implement the \textsc{MultiNest} algorithm \cite{FeHo,FeHB,FHCP}.} and, in fact, there is no sharp peak in the value distribution of the extra parameters in the measure, including the amplitudes $A$ and $B$ in \Eq{heurlog}. However, certain regions in the parameter space are excluded by our analysis due to the fact that CMB data disfavour log oscillations. These regions are compatible with the independent constraints found in Standard-Model and astrophysical observations \cite{frc12,qGW}. From the constraint on the fractional exponent $\a$, we will find that the Hausdorff dimension of space in the ultraviolet (UV) cannot exceed
\be\label{dhst}
\boxd{\begin{matrix}
N=2:\qquad & \dh^{\rm \,space}\lesssim 0.3\qquad\text{(UV)}\,,\\
N=3:\qquad & \dh^{\rm \,space}\lesssim 1.9\qquad\text{(UV)}\,,\\
N=4:\qquad & \dh^{\rm \,space}\lesssim 1.7\qquad\text{(UV)}\,,
\end{matrix}}
\ee
where the parameter $N$ will be introduced in the next section (roughly speaking, it is the number of ``copies'' a self-similar spacetime is described with at each level of iteration).

In section \ref{setup}, we review the multi-fractional model with $q$-derivatives. After calculating the frequency distribution for a black body in multi-fractional spacetimes with $q$- and weighted derivatives, constraints from the CMB black-body spectrum using FIRAS data are presented in section \ref{firas} and compared with the bounds found in dimensional regularization. We will also compare the constraints obtained for the multi-fractional theory with $q$-derivatives with the one with weighted derivatives (e.g., \cite{frc11}) and with the toy model leading to \Eq{boue3}. In section \ref{infl}, we focus our attention on the theory with $q$-derivatives and write the inflationary observables which will be used in the numerical analysis of section \ref{num}. Section \ref{disc} is dedicated to a discussion and comparison with previous bounds.


\section{Setup}\label{setup}

Most of this paper will focus on the multi-fractional theory with $q$-derivatives. The latter is defined in an apparently simple-minded way (it is, perhaps, the simplest of the four multi-fractional theories \cite{frc11}): one picks their favorite field-theory or gravitational covariant action and replaces everywhere coordinates $x^\mu$ ($\mu=0,1,\cdots,D-1$) with a certain fixed profile $q^\mu(x^\mu)$ we will define below according to the rules of fractal geometry. The measure $\rmd^Dx\to \rmd q^0(t)\,\rmd q^1(x^1)\dots \rmd q^{D-1}(x^{D-1})$ is therefore factorizable in the coordinates. The profiles $q^\mu(x^\mu)$ depend on some fundamental scales which introduce a crucial difference with respect to general relativity. In fact, the map $x^\mu\to q^\mu(x^\mu)$ is not a trivial coordinate change because, according to the multi-scale paradigm, one must specify measurement units attached to the coordinates. In other words, one must choose a coordinate frame where physical measurements take place. By definition of the theory, this frame is spanned by the scale-independent $x^\mu$. The intuitive reason is that the scale of our clocks and rods is independent of the intrinsic scale of the phenomena detected: they are defined a priori according to some conventions. On the other hand, the $q^\mu(x^\mu)$ would correspond to adaptive but non-physical clocks and rods. A longer discussion can be found in \cite{frc11,trtls} but we will provide new arguments about the frame choice that will apply to the cosmological context (see below \Eq{rhoev}).

For simplicity, we will consider identical profiles $q^\mu(x^\mu)=q(x^\mu)$ which may differ only in the value of the parameters therein.


\subsection{Action, measure and momentum space}

The gravitational and cosmological dynamics of the multi-fractional theory with $q$-derivatives has been developed in \cite{frc11}. The action in $D$ dimensions is given by 
\be\label{Sgq}
S =\frac{1}{2\kappa^2}\int\rmd^Dx\,v\,\sqrt{-g}\,({}^q R-2\Lambda)+S_{\rm m}\,,
\ee
where $\k^2=8\pi G$, $g$ is the determinant of the metric $g_{\mu\nu}$ and $S_{\rm m}$ is the matter action (possibly including the inflaton). Apart from the metric structure, the geometry also has a pre-fixed multi-scale structure embodied in the factorizable measure weight $v(x)=v(t)v(x^1)\dots v(^{D-1})$. In general, along each direction $x$, one can define an effective geometric coordinate
\be\label{geom}
q(t)=\int^t\rmd t'\,v(t')\,,\qquad q(x^i)=\int^{x^i}\rmd {x'}^i\,v({x'}^i)\,,
\ee
where $i=1,\dots,D-1$ denotes spatial coordinates. The curvature invariant in \Eq{Sgq} is the Ricci scalar defined in terms of geometric coordinates, ${}^q R=g^{\mu\nu}{}^q R^\s_{~\mu\s\nu}$, where
\ba
{}^q R^\rho_{~\mu\s\nu} &:=& \frac{1}{v_\s}\p_\s {}^q\G^\rho_{\mu\nu}-\frac{1}{v_\nu}\p_\nu {}^q\G^\rho_{\mu\s}+{}^q\G^\tau_{\mu\nu}\,{}^q\G^\rho_{\s\tau}-{}^q\G^\tau_{\mu\s}\,{}^q\G^\rho_{\nu\tau}\,,\label{riemq}\\
{}^q\G^\rho_{\mu\nu} &:=& \tfrac12 g^{\rho\s}\left(\frac{1}{v_\mu}\p_{\mu} g_{\nu\s}+\frac{1}{v_\nu}\p_{\nu} g_{\mu\s}-\frac{1}{v_\s}\p_\s g_{\mu\nu}\right)\,.\label{leciq}
\ea

The profile $v(x)$ is dictated by symmetry considerations in the multi-fractal geometry \cite{fra4,frc1,frc2}. One can show that a simple but fairly complete Ansatz encoding all the main properties of both deterministic and random multi-fractals is
\bs\label{logos}\ba
q(t) &=& t+\frac{t_*}{\a_0}\left(\frac{t}{t_*}\right)^{\a_0} F_{\om_0}(\ln t)\,,\\
F_{\om_0}(\ln t) &=& 1+A_0\cos\left(\om_0\ln\frac{t}{t_\infty}\right)+B_0\sin\left(\om_0\ln\frac{t}{t_\infty}\right)\,,
\ea\es
where $0<\a_0<1$ is called fractional exponent. Here we assume $t\geq 0$ for the sake of simplicity. In particular,
\be
v(t)=1+\left(\frac{t}{t_*}\right)^{\a_0-1}\left[1+A_+^0\cos\left(\om_0\ln\frac{t}{t_\infty}\right)+B_-^0\sin\left(\om_0\ln\frac{t}{t_\infty}\right)\right]\,,
\ee
where
\be
A_\pm^0=A_0\pm\frac{B_0\om_0}{\a_0}\,,\qquad B_\pm^0=B_0\pm\frac{A_0\om_0}{\a_0}\,.
\ee
Similar expressions hold for the other directions but with absolute values:
\bs\ba
q(x^i) &=& x^i+\frac{\ell_*}{\a}\left|\frac{x^i}{\ell_*}\right|^\a F_\om(\ln|x^i|)\,,\\
F_\om(\ln |x^i|) &=& 1+A\cos\left(\om\ln\left|\frac{x^i}{\ell_\infty}\right|\right)+B\sin\left( \om\ln\left|\frac{x^i}{\ell_\infty}\right| \right)\,.
\ea\es
The exponent $0<\a<1$ and the frequency $\om>0$ have been chosen to be the same for all spatial directions.

Such geometries are not a trivial rewriting of standard ones because measurement units are assigned to coordinates a priori. In particular, the results of experiments are \emph{defined} to be cast in terms of the coordinates $x^\mu$, which have length units, $[x^\mu]=-1$. There is neither Poincar\'e nor Lorentz invariance with respect to these coordinates, but the measure in the action \Eq{Sgq} is defined so that to be invariant under the non-linear transformations $q({x'}^\mu)=\Lambda_\nu^{\ \mu}q(x^\nu)+Q^\mu$.

Logarithmic oscillations are typical of deterministic fractals \cite{frc2,NLM}. They encode a discrete dilation symmetry under which the oscillatory part of the action is invariant (discrete scale invariance, DSI \cite{Sor98}). For spatial directions,
\be\label{dsira}
x^i\,\to\, \la_\om^m x^i\,,\qquad \la_\om=\exp\left(-\frac{2\pi}{\om}\right)\,,
\ee
where $m$ is integer. All deterministic fractals possess a DSI. A typical example is the middle-third Cantor set: at each iteration, one takes $N=2$ smaller copies of the set identical to larger ones under a rescaling ratio $\la_\om=1/3$. The capacity (equal to the Hausdorff dimension in this case) is then $d_\textsc{c}=-\ln N/\ln\la_\om$. In the context of gravity, a DSI signals a \emph{discrete} structure of spacetime at ultra-microscopic scales of order of $t_\infty$, $\ell_\infty$, \emph{even if the theory is embedded in a continuum}. At large scales, oscillations are averaged out to an effective continuum. The scales $t_\infty$, $\ell_\infty$ entering the log oscillations have been identified with, respectively, the Planck time $\tp$ and the Planck length $\lp$ thanks to a theoretical connection between multi-fractional and non-commutative spacetimes \cite{ACOS}. However, they can both be left unspecified. In general, however, they will be smaller than the other characteristic scales entering the system ($t_*$, $\ell_*$) via the power-law function in \Eq{logos}. These scales mark the transition between a macroscopic phase where geometry is standard and a phase where geometry becomes anomalous and the dimensionality of spacetime differs from the topological dimension $D$. At length scales below those of the second phase, the discrete structure emerges.

The parameter space of the theory is not continuous in the direction of the frequency $\om$, i.e., $\om$ takes discrete values. In fact, the structure of the geometry at scales $\sim \ell_\infty$ is that of a deterministic fractal made of $N$ copies rescaled by the ratio $\la_\om$ in \Eq{dsira}. To determine the number $N$, we recall that the Hausdorff dimension is $\a$ along each direction and that the Hausdorff dimension and the capacity coincide for this type of geometries, $\dh=d_\textsc{c}$. Therefore, from $\a=-\ln N/\ln\la_\om=\om\ln N/(2\pi)$ we have \cite{trtls}
\be\label{omspe}
\om=\om_N:=\frac{2\pi\a}{\ln N}\,.
\ee
For example, when $\a=1/2$ and $N=2,3,\dots$, we have $\la_\om=1/N^2$ and
\ba
&& N=2:\qquad \om=\om_2\approx 4.53\,,\qquad \la_\om=\tfrac14\,,\nonumber\\
&& N=3:\qquad \om=\om_3\approx 2.86\,,\qquad \la_\om=\tfrac19\,,\nonumber\\
&& N=4:\qquad \om=\om_4\approx 2.27\,,\qquad \la_\om=\tfrac{1}{16}\,,\nonumber\\
&& \,\,\,\quad\vdots\label{omspe2}\\
&& N=10:\qquad \om=\om_{10}\approx 1.36\,,\qquad \la_\om=\tfrac{1}{100}\,.\nonumber\\
&& \,\,\,\quad\vdots\nonumber
\ea
In the last part of the paper, we will take into account this discreteness of values. The spectrum $\om_0:=2\pi\a_0/\ln N$ in the time direction is similar to the above but it will not be needed here.

The structure of momentum space reflects the same features of position space. For each direction and assuming one characteristic scale for all $\mu$, the geometric coordinate is\footnote{A typo in (6.35) and (6.36) of \cite{frc11} is here corrected.}
\ba
p^0(E,E_*) &:=& \frac{1}{q\left(\frac{1}{E},\frac{1}{E_*}\right)}\label{pik10}\\
&=&\frac{E}{1+\frac{1}{\a_0}\left|\frac{E_*}{E}\right|^{\a_0-1} F_{\om_0}(-\ln|E|)}\label{pik20}\,,
\ea
and
\ba
p(k^i,E_*) &:=& \frac{1}{q\left(\frac{1}{k^i},\frac{1}{E_*}\right)}\label{pik1}\\
&=&\frac{k^i}{1+\frac{1}{\a}\left|\frac{E_*}{k^i}\right|^{\a-1} F_\om(-\ln|k^i|)}\label{pik2}\,,
\ea
where $E=k^0$. These formul\ae\ \cite{frc11} follow from some minimal requirement on the structure of momentum space of the theory \cite{frc1,frc3}, namely, the existence of an invertible Fourier transform at all scales. It can be easily shown that $pq\simeq kx$ at large scales, while $pq\simeq (kx)^\a$ at mesoscopic scales larger than the frequency of log oscillations.


\subsection{Cosmological dynamics}

The Einstein equations stemming from the dynamics \Eq{Sgq} are given by
\be\label{eeq}
{}^qR_{\mu\nu}-\frac12 g_{\mu\nu} ({}^q R-2\Lambda)=\k^2\, {}^qT_{\mu\nu}\,.
\ee
The Friedmann and continuity equation for a perfect fluid on a Friedmann--Lema\^itre--Robertson--Walker (FLRW) are 
\ba
&&\left(\frac{D}{2}-1\right) \frac{H^2}{v^2}=\frac{\kappa^2}{D-1}\,\rho+\frac{\Lambda}{D-1}-\frac{\textsc{k}}{a^2}\,,\label{fri}\\
&& \dot\rho+(D-1)H(\rho+P)=0\,,\label{ra}
\ea
where $\textsc{k}=0,\pm1$ is the intrinsic curvature, 
$H$ is the Hubble parameter, and a dot represents 
a derivative with respect to the cosmic time $t$.
For a power-law expansion in the geometric coordinate,
\be\label{plexp}
a(t)=[q(t)]^p\,,
\ee
one has
\be\label{hup}
H=p\frac{\dot q(t)}{q(t)}=p\frac{v(t)}{q(t)}\,.
\ee
The universe does not expand monotonically at early times and undergoes an infinite series of contractions and expansion. This evolution can be studied analytically \cite{frc11}.


\subsection{Proper and geometric energy density}

The quantities $\rho$ and $P$ in equations \Eq{fri} and \Eq{ra} are, respectively, the \emph{geometric} energy density and pressure of a perfect fluid. It is important to stress the difference between geometric and proper energy density; this point was not touched upon in \cite{frc11} but it will play a crucial role in section \ref{firas}. 

Let $E$ be the eigenvalue of a generic quantum matter Hamiltonian $\hat H$ defined on a proper three-dimensional spatial volume $V$. In standard spacetimes, such eigenvalue is interpreted as the energy associated with the matter content in $V$. Classically, the proper energy density is then 
\be
\varrho:=\frac{E}{V}\,.
\ee
On the other hand, in multi-fractional theories there are two possible interpretations of the eigenvalue $\cE$ of the quantum Hamiltonian operator $\hat H$ defined on the geometric volume $\cV$. By construction, $\cE$ has engineering dimension $[\cE]=1$, i.e., an energy.\footnote{This is obvious in the theory with $q$-derivatives, since $[q]=-1$ and all equations are formally the usual ones. For instance, for a massive non-relativistic particle $\hat H\propto \p_q^2/m+\dots\,$, where $[m]=1$.} Then, $\cE$ can be identified either with the actual energy of a mode (as done in the theory with weighted derivatives \cite{frc5}) or with the geometric energy $\cE=p^0(E)$, where $E$ is the proper energy. However, in order to have a non-trivial time evolution $\cE(E)$ must be conjugate to $q(t)$ in the sense specified by \Eq{pik1}. The geometric energy density is thus
\be
\rho:=\frac{\cE(E)}{\cV(V)}\,.
\ee
The relation $\cV(V)$ between the proper and the geometric volume is highly non-trivial and, in general, it cannot even be defined as a one-to-one correspondence. However, it can be simplified for a cubic region $V=L^3$: in this case, $\cV:=[q(L)]^3=[q(V^{1/3})]^3$. This expression is meaningful even when it cannot be inverted, as it happens in multi-fractional spacetimes with log oscillations ($q$ is a transcendental function). But even when $q$ and $p$ can be inverted (for instance, for multi-fractional measures with no oscillations), one cannot express $\rho$ as a function of $\varrho$, since
\be\label{rhoev}
\rho(E,V)\neq \rho(E/V)=\rho(\varrho)\,.
\ee
In a cosmological setting, this obstructions fixes unambiguously the energy density which is actually measured. For consistency with the cosmological principle, it must be $\rho$, the geometric energy density appearing in the Friedmann equations \Eq{fri} and \Eq{ra}. In fact, by construction $\rho=\rho(t)$ but if $\rho$ is homogeneous then $\varrho$ cannot be homogeneous, too, due to the inequality \Eq{rhoev}. In other words, the explicit coordinate dependence of the spatial measure forbids to implement translation invariance in both (proper and geometric) coordinate frames. As a consequence, physical measurements will involve geometric rather than proper energy densities.



\section{CMB black-body spectrum in multi-fractional spacetimes}\label{firas}

The spectrum of the cosmic microwave background is a black-body spectrum with temperature $T\approx 2.725\,\mbox{K}$ to a high degree of precision \cite{Fix96}. This can give us interesting information on how far away from three spatial dimensions could the universe be at the epoch of radiation-matter decoupling. First, we need to compute the black-body spectrum in multi-fractional geometries with $q$-derivatives. In this section only, we do not use natural units; $hc\approx 1.239\times 10^{-6}\,{\rm eV}\cdot {\rm m}$ will indicate the non-reduced Planck's constant times the speed of light.


\subsection{Theory with \texorpdfstring{$q$}{}-derivatives}\label{qth}

Consider a thermal bath of photons in a box of linear size $L$. The energy levels of a mode of the electromagnetic field in the cavity are the eigenvalues $\cE_l=(l+1/2)\ve$ of the quantum Hamiltonian, where $l=0,1,\dots$ is the occupation number of photons in the mode. According to the mapping $k^\mu\to p^\mu(k^\mu)$, $\ve=p^0(\bar E)=hcn/[2q(L)]$ is the geometric energy of a single photon, where $\bar E$ is the actual energy of a photon.\footnote{This is the multi-fractional analogue of the usual counting of states in a box, where the energy of a photon is inversely proportional to the box size: $\bar E=hcn/(2L)$.} As the second equality shows, $\ve$ depends on the size of the box and on the wave-length of the mode (parametrized by the integer $n$). 

Formally, the probability distribution of the mode over the energy levels is the same as usual: $P_l=\rme^{-\b \cE_l}/Z(\b)$, where $Z(\b):=\sum_{l=0}^{+\infty}\rme^{-\b \cE_l}=\rme^{-\b \cE_l/2}/(1-\rme^{-\b \cE_l})$ is the partition function of the mode. To recognize correctly the relation between $\b$ and the temperature of the thermal bath of photons in the cavity (in the cosmological context, the CMB temperature), we recall that the variable conjugate to the geometric momentum $p(k)$ is $q(x)$, so that $p(k)\,q(x)\sim f(kx)$ at all scales for polynomial functions. Extending this requirement to generic functions, one reaches formula \Eq{pik1}, $p(k)\,q(1/k)=1$. In the standard case, $p\,\dot{=}\,\mathbbm{1}\,\dot{=}\,q$ (here $\dot{=}$ indicates an equality between functionals) and this relation is trivial, $k\times(1/k)\sim 1$. Imposing the same relation for $\cE$ and its conjugate $\b(T)$, one finds that $\cE\,\dot{=}\,\mathbbm{1}\,\dot{=}\,\b$ in the standard case and $\b\,\dot{=}\,q$ in the multi-fractional case. Therefore,
\be
\b(T)=q\left(\frac{1}{k_{\rm B} T}\right)=\frac{1}{k_{\rm B} T}\left[1+\frac{1}{\a_0}\left(\frac{T}{T_*}\right)^{1-\a_0} F_{\om_0}(-\ln|k_{\rm B} T|)\right]\,,
\ee
where $k_{\rm B}\approx 8.617\times 10^{-5}\,{\rm eV}/{\rm K}$ is Boltzmann's constant and $T_*=E_*/k_{\rm B}$.

The thermodynamical average energy of a mode is $\langle\cE\rangle:=-\rmd\ln Z/\rmd\b=\ve/2+\ve/(\rme^{\b\ve}-1)$. The first contribution is the vacuum energy, while the second defines the geometric energy $\cU(U)=p^0(U)$ (with $U$ being the proper energy) of an infinite box according to
\be\label{Ust}
\cU=\int_0^{+\infty}\rmd\ve\,\mu(\ve)\,\frac{\ve}{\rme^{\b\ve}-1}\,,
\ee
where $\mu(\ve)$ is the density of states. This quantity is calculated by taking twice (for the two photon states) one eight (the positive octant) of a spherical volume in the state space, which is a geometric spherical sector in the multi-fractional case: $\mu(\ve)\,\rmd\ve=2(1/8)4\pi n^2\,\rmd n=8\pi [q(L)/(hc)]^3\ve^2\,\rmd\ve$. Overall, the geometric energy density in the cavity is
\be
\rho_{\rm r}:=\frac{\cU}{[q(L)]^3}=\frac{8\pi}{(hc)^3}\int_0^{+\infty}\rmd\ve\,\frac{\ve^3}{\rme^{\b\ve}-1}\,.
\ee
The frequency is the number of occurrences of events per time unit. It is an operational quantity directly related to experiments, which means that it must be related to the actual energy rather than the geometric one. Therefore, we write $E=:h\nu$. In terms of the frequency,
\be\label{bbsms}
\rho_{\rm r}=\frac{4\pi}{c}\int_0^{+\infty}\rmd\Om(\nu)\,\cB_\nu(T)\,,\qquad \cB_\nu(T)=\frac{2h}{c^2}\frac{\Om^3(\nu)}{\rme^{\b(T)h\Om(\nu)}-1}\,,
\ee
where
\be
\Om(\nu):=\frac{\ve(h\nu)}{h} =\frac{\nu}{1+\frac{1}{\a_0}\left(\frac{\nu}{\nu_*}\right)^{1-\a_0} F_{\om_0}(-\ln\nu)}\label{pnu}\,,\qquad \nu_*=\frac{E_*}{h}.
\ee
The measured spectrum is the integrand in \Eq{bbsms} with measure $\rmd\nu$, i.e.,
\be
B_\nu(T)=\Om'(\nu)\,\cB_\nu(T)\,.
\ee
In the standard case, $\Om=\nu$ and
\be\label{Bstan}
B_\nu(T)\to \cB_\nu(T)\to\frac{2h}{c^2}\frac{\nu^3}{\rme^{h\nu/(k_{\rm B}T)}-1}\,.
\ee

Integrating \Eq{bbsms}, one obtains the temperature law for the radiation geometric energy density, 
\be
\rho_{\rm r}=\frac{8\pi^5}{15(hc)^3}\frac{1}{\b^4(T)}\,.
\ee
For $T_*\gg T$, $\rho_{\rm r}\simeq [8\pi^5 k_{\rm B}^4/(15h^3c^3)] T^4[1-(4/\a_0)(T/T_*)^{1-\a_0} F_{\om_0}(-\ln T)]$ and the geometric energy density is slightly smaller than the proper energy density $\varrho_{\rm r}=U/L^3\propto T^4$ in the standard case. For $T_*\ll T$, $\rho_{\rm r}\simeq [8\pi^5(\a_0 k_{\rm B}T_*)^4/(15h^3c^3)](T/T_*)^{4\a_0}/F_{\om_0}^4$ and the energy density has a milder dependence on the temperature.

To fit the \textsc{cobe} data \cite{Fix96,cobe} with the spectrum \Eq{bbsms}, we ignore oscillations, $F_\om=1$. This is a reasonable preliminary approximation, since we expect that the logarithmic modulation of the spectrum is a secondary effect with respect to the change of the spectral index. Also, letting $\a_0$ free would always lead to a best fit with $\a_0\approx 1$. This is expected, since the frequency distribution is very close to the standard three-dimensional distribution \Eq{Bstan} and it is more likely to reproduce it with $\a_0=1$ and \emph{any} $\nu_*$ rather than with $\a_0\neq 1$ and some~$\nu_*$.

Let $B_\nu=C\Om'(\nu)\Om^3(\nu)/[\rme^{\b(T)h\Om(\nu)}-1]$ be the fit function, where $\Om'(\nu)=\a_0^2[1+(\nu/\nu_*)^{1-\a_0}]/[\a_0+(\nu/\nu_*)^{1-\a_0}]^2$. The normalization $C$ will be determined numerically together with the other parameters but it is physically unimportant. The method we employ is quite simple: we define the best fit as the function minimizing the square of the residuals $|B_\nu^{\rm obs}-B_\nu|$.

\begin{figure}
\centering
\includegraphics[width=10cm]{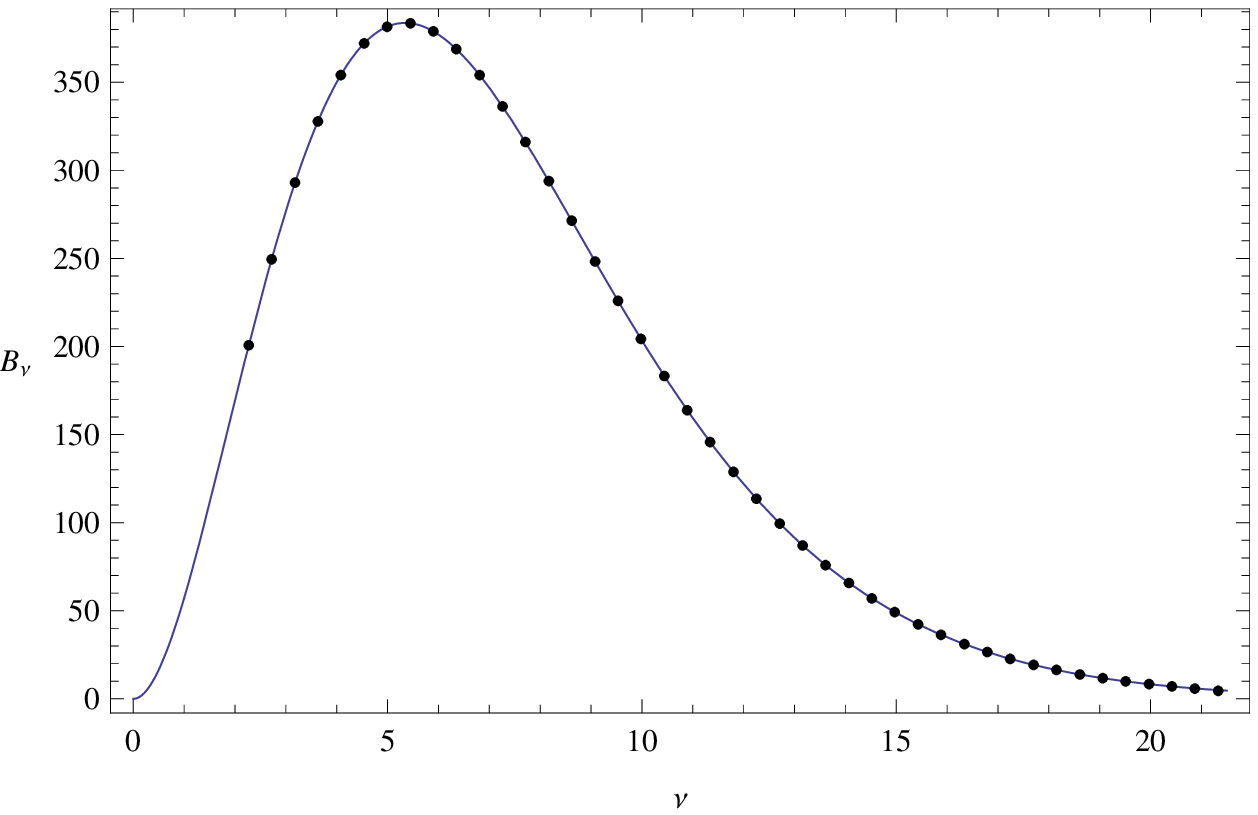}
\caption{\label{fig1} The best fit \Eq{bfit} of FIRAS data \cite{cobe} with the multi-fractional black-body spectrum \Eq{bbsms}.}
\end{figure}

For $\a_0=1/2$, figure \ref{fig1} shows the best fit to the data, with $C\approx 39.73$ and
\be
\label{bfit}
\nu_* \approx 4.5 \times 10^{14}\,{\rm cm}^{-1}\qquad\Rightarrow\qquad E_*^\text{best-fit}\approx 56\,\mbox{GeV}\,,
\ee
where the propagated error is many orders of magnitude smaller than the quoted values. For $\a_0=1/4$,
\be\label{bfit2}
\nu_* \approx 1.5 \times 10^{12}\,{\rm cm}^{-1}\qquad\Rightarrow\qquad E_*^\text{best-fit}\approx 180\,\mbox{MeV}\,.
\ee

Here and in all the following examples,  the black-body temperature is obtained by inverting the relation $\b(T)=\b_\text{best-fit}$ with $T_*=h\nu_*/k_{\rm B}$. It equals $T\approx 2.725\,\mbox{K}$ within the numerical and experimental uncertainty. 

These best fits place a lower bound on the characteristic energy scale of the theory: any actual $E_*$ higher than $E_*^\text{best-fit}$ will be in agreement with the data:
\be
E_*> E_*^\text{best-fit}\sim 1-100\,\mbox{GeV}\,.
\ee
This lower bound is not very stringent: in reality we would expect an energy scale higher (or even much higher) than present particle-physics experiments, $E_*>1\,\mbox{TeV}$. However, such low values are not surprising, since the characteristic energy scale of the CMB black-body spectrum is quite low,
\be
E_\text{CMB}=k_{\rm B}T
\approx 2.3\times 10^{-4}\,\mbox{eV}\,.
\ee
Energies of the order of \Eq{bfit} or \Eq{bfit2} are much higher than $E_\text{CMB}$ and multi-scale corrections are always negligible. We can conclude that the information on the black-body spectrum is insufficient to place interesting constraints on the characteristic energy scale of the theory. However, such information is quite competitive with the bounds obtained in particle physics, as we will see in section \ref{disc}. To make a full comparison in the case $\a_0=1/2$, we convert the best fit \Eq{bfit} into bounds on the time and length scales of the measure in position space, using Planck units as argued in \cite{frc12,frc13}:
\ba
t_*^\text{best-fit} &:=& t_\Pl \frac{m_\Pl}{E_*^\text{best-fit}} \approx 1.2\times 10^{-26}\,{\rm s}\,,\label{tbf1}\\
\ell_*^\text{best-fit} &:=& \ell_\Pl \frac{m_\Pl}{E_*^\text{best-fit}} \approx 3.5\times 10^{-18}\,{\rm m}\,.\label{lbf1}
\ea


\subsection{Theory with weighted derivatives}

The theory with $q$-derivatives is only one among the four multi-fractional scenarios proposed so far (see \cite{frc11} for a summary and references) but it is the only one where, thanks to the mappings $x\to q(x)$ and $k\to p(k)$ from standard general relativity, the inflationary spectra have been fully worked out. For this reason, we will compute the likelihood distribution of the inflationary parameters of the model only for this case. However, it is both easy and interesting to extend the discussion of section \ref{qth} to the theory with weighted derivatives, in order to compare the upper bounds on $E_*$. Although it is already known that the evolution of the universe is qualitatively different in all these models \cite{fra2,frc11}, their scale hierarchies have never been confronted with one another. We cannot say much about the theory with normal derivatives, since in that case there is no known well-defined invertible momentum transform \cite{frc3}.

For weighted derivatives, momentum space is endowed with a measure $\rmd^Dk\,w(k)$, where $w(k)$ is some factorized function of the $D$ components of $k^\mu$. However, there are no geometric coordinates entering the Fourier transform, so that $\cE_l=E_l$, $\ve=\bar E$ and $\b=1/(k_{\rm B} T)$ in all formul\ae\ in the black-body calculation. However, the density of states is weighted with $w$ \cite{frc7}, so that equation \Eq{Ust} is replaced by
\be\nonumber
U=\int_0^{+\infty}\rmd\ve\,w(\bar E)\,\mu(\bar E)\,\frac{\bar E}{\rme^{\b\bar E}-1}\,,\qquad   \mu(\ve)=8\pi \left(\frac{L}{hc}\right)^3\bar E^2\,.
\ee
The multi-scale Planck law eventually reads
\be
B_\nu(T)=\frac{2h}{c^2}\frac{\nu^3w(\nu)}{\rme^{h\nu/(k_{\rm B} T)}-1}\,.
\ee
Contrary to the theory with $q$-derivatives, there is no principle fixing the form of $w(\nu)$. For definiteness, however, we can take it to be $v(1/\nu)$. 
Ignoring log oscillations, we have
\be
w(\nu)=\left[1+\left(\frac{\nu}{\nu_*}\right)^{1-\a_0}\right]^3\,.
\ee
For $\a_0=1/2$, the best fit obtained by minimizing the square of the residuals yields (again, errors are negligible)
\be
\nu_*\approx 3.0\times 10^{9}\,{\rm cm}^{-1}\qquad\Rightarrow\qquad E_*^\text{best-fit}\approx 374\,\mbox{keV}\,,
\ee
corresponding to the time and length scales
\ba
t_*^\text{best-fit} &=& t_\Pl \frac{m_\Pl}{E_*^\text{best-fit}} \approx 1.8\times 10^{-21}\,{\rm s}\,,\label{tbf2}\\
\ell_*^\text{best-fit} &=& \ell_\Pl \frac{m_\Pl}{E_*^\text{best-fit}} \approx 5.3\times 10^{-13}\,{\rm m}\,.\label{lbf2}
\ea
For $\a_0=1/4$, 
\be
\nu_*\approx 3.4\times 10^6\,{\rm cm}^{-1}\qquad\Rightarrow\qquad E_*^\text{best-fit}\approx 417\,\mbox{eV}\,.
\ee
With respect to the case with $q$-derivatives, these bounds are weaker.

We can compare these constraints with those obtained in toy models of dimensional regularization, where the spacetime dimensionality is fixed at all scales \cite{CO}. In that case, there is a strong bound on the deviation from three dimensions. The black-body spectral distribution in $D-1=3+\e$ spatial dimensions is
\be\label{Bdr}
B_\nu^{(\e)}(T)=C \frac{\nu^{3+\e}}{\rme^{h\nu/(k_{\rm B}T)}-1}\,,
\ee
for which we found the best fit
\be
\e\approx (-5.3\pm 10.6)\times 10^{-5}\,,\qquad C\approx 39.7\,,
\ee
in agreement with the estimate made in \cite{CO}.

This model is equivalent to a fractional model with weighted derivatives with $v(t)=1$ and $v=\prod_{i=1}^3v(x^i)\propto \prod_i|x^i|^{\a-1}$, so that $3+\e=D-1=3\a$ and $\a-1=\e/3$. Since there is no non-trivial hierarchy of scales, fractional geometries are strongly constrained by various experiments and they do not constitute realistic models of physics \cite{frc1,frc2}. As we have seen, as soon as we introduce a characteristic scale $E_*$ the fractional index $\a$ can take any theoretically allowed value.


\section{Inflationary observables}\label{infl}

In \cite{frc11}, some qualitative features of inflationary power spectra in multi-fractional theory with $q$-derivatives where sketched. Here we present a more rigorous analysis and a comparison with experiments.

In geometric coordinates $q$ and $p$, the dynamical equations take the usual form by definition of the theory. As a consequence, it is not necessary to present an explicit calculation of the dynamics of perturbations, since it follows exactly the same steps as in general relativity (see, e.g., \cite{MFB}), the only difference been the replacements $x^\mu\to q^\mu(x^\mu)$ and $k^\mu\to p^\mu(k^\mu)$. In particular, the decomposition of the metric into scalar, tensor and vector modes is the same as usual. Assuming that the only matter content is a scalar field $\phi$, one can define a gauge-invariant curvature perturbation $u=z\cR$, where $z=a\dot\phi/H$ is a homogeneous background function and $\cR=(H/\dot\phi)\de\phi+\Psi$ is the curvature perturbation on comoving hypersurfaces, made of a combination of scalar fluctuations in the inflaton and in the metric. Then, the Mukhanov--Sasaki equation in momentum space and in geometric conformal time becomes
\be\label{mukeq}
u_k''+\left(\tilde k^2-m_{\rm eff}^2\right)u_k=0\,,
\ee
where, apart from the functional difference in the effective mass term $m_{\rm eff}^2\simeq(aH/v)^2+\dots$, the absolute value $k=|{\bf k}|$ of the spatial comoving wave-number is replaced by
\be\label{tik}
\tilde k:=\sqrt{\sum_i p^2(k^i)}=\sqrt{k^2+\dots}\,\,.
\ee

The power spectra can be constrained independently of the details of the inflationary dynamics by writing them as power laws in $q$-momentum space:
\be\label{powspe}
\cP_{\rm s,t}=\cA_{\rm s,t}\left(\frac{\tilde k}{\tilde k_0}\right)^{n}\,,\qquad n=n_{\rm s}-1,\,n_{\rm t}\,,
\ee
where $\tilde k_0=\tilde k(k_0)$ is the pivot scale of the experiment.

Assuming that $k^1=k^2=k^3=k/\sqrt{3}$ and rescaling the characteristic energy scales as $k_*:=\sqrt{3}E_*$ and $k_\infty:=\sqrt{3}E_\infty$, we have
\be
\tilde k=p(k,k_*)\,.
\ee 
Expanding \Eq{powspe} for small oscillation amplitudes $A,B\ll 1$, one obtains the main equation of this work:\footnote{We can compare our spectrum \Eq{Pmusc} with the fit function (79) used by the \textsc{Planck} collaboration and coming from different theoretical models \cite{P1520}. The first difference is that our $\cP(k)$ is of the form ``(power-law with two regimes)$\times$($1+$cos$+$sin),'' while eq.\ (79) of \cite{P1520} is (after expanding the cosine and absorbing the constant phase into the amplitudes) of the form ``(standard power-law)$\times$($1+$cos$+$sin).'' Second, the ratio in the logarithm is $k_\infty/k$ in the multi-scale case and $k/k_0$ in the \textsc{Planck} case.}
\ba
\cP(k) &\simeq&\cA_{\rm s,t}\left(\frac{k}{k_0}\frac{\a+\left|\frac{k_0}{k_*}\right|^{1-\a}}{\a+\left|\frac{k}{k_*}\right|^{1-\a}}\right)^n\left[1+ A n\cos\left(\om\ln\frac{k_\infty}{k}\right)+B n\sin\left(\om\ln\frac{k_\infty}{k}\right)\right.\nonumber\\
&&\qquad\qquad\qquad\qquad\qquad \left.
~~-A n\cos\left(\om\ln\frac{k_\infty}{k_0}\right)-B n\sin\left(\om\ln\frac{k_\infty}{k_0}\right)\right],\label{Pmusc}
\ea
where $k_\infty$ is the wave-number at which discreteness effects become important. For $k\ll k_*< k_\infty$, one recovers the usual power spectrum $\cP(k) \propto k^n$, and so one does when $k_*\ll k< k_\infty$, but with an effective index
\be\label{tilden}
\tilde n=\a n\,.
\ee
At these scales, one can obtain an almost scale-invariant spectrum even for models where $n\not\ll 1$, provided $\a$ is sufficiently small. The possibility to have almost scale invariance away from the slow-roll regime is typical of models with dimensional flow, including rainbow gravity \cite{AAGM} or varying-speed-of-light scenarios. The change of dimensionality affects the dynamics of the Hubble horizon in a way that solves the problems of the hot big bang and produces large-scale constant perturbations.

\begin{figure}
\centering
\includegraphics[width=10cm]{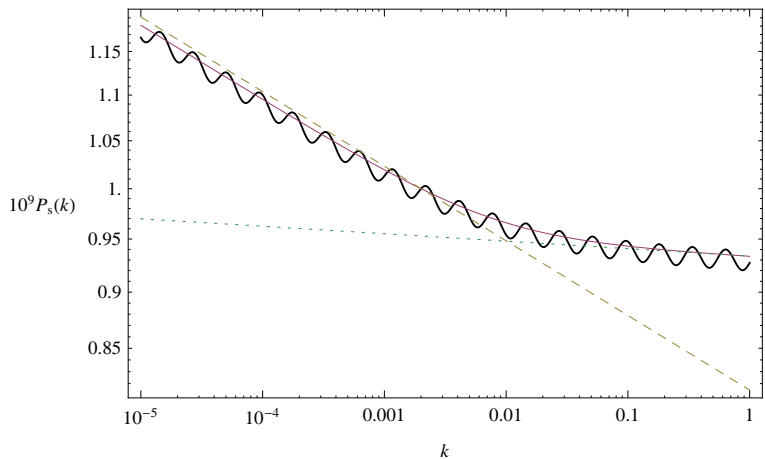}
\caption{\label{fig2} Solid thick curve: the scalar spectrum \Eq{Pmusc} for $\cA_{\rm s}=10^{-9}$, $k_0=0.002\,\text{Mpc}^{-1}$, $\a=0.1$, $k_*=0.1\,\text{Mpc}^{-1}$, $n_{\rm s}=0.967$, $A=0.3$, $B=0$ and $\om=10$ (here, $\om$ is large for illustrative purposes only). Solid thin curve: the same spectrum but without log oscillations ($A=0$). Dashed line: the standard power spectrum $\cA_{\rm s}(k/k_0)^{n_{\rm s}-1}$. Dotted line: the power spectrum at small scales, $\cP_{\rm s}\simeq\cA_{\rm s}(k/k_0)^{\a(n_{\rm s}-1)}$.}
\end{figure}

\begin{figure}[ht]
\centering
\includegraphics[width=7.7cm]{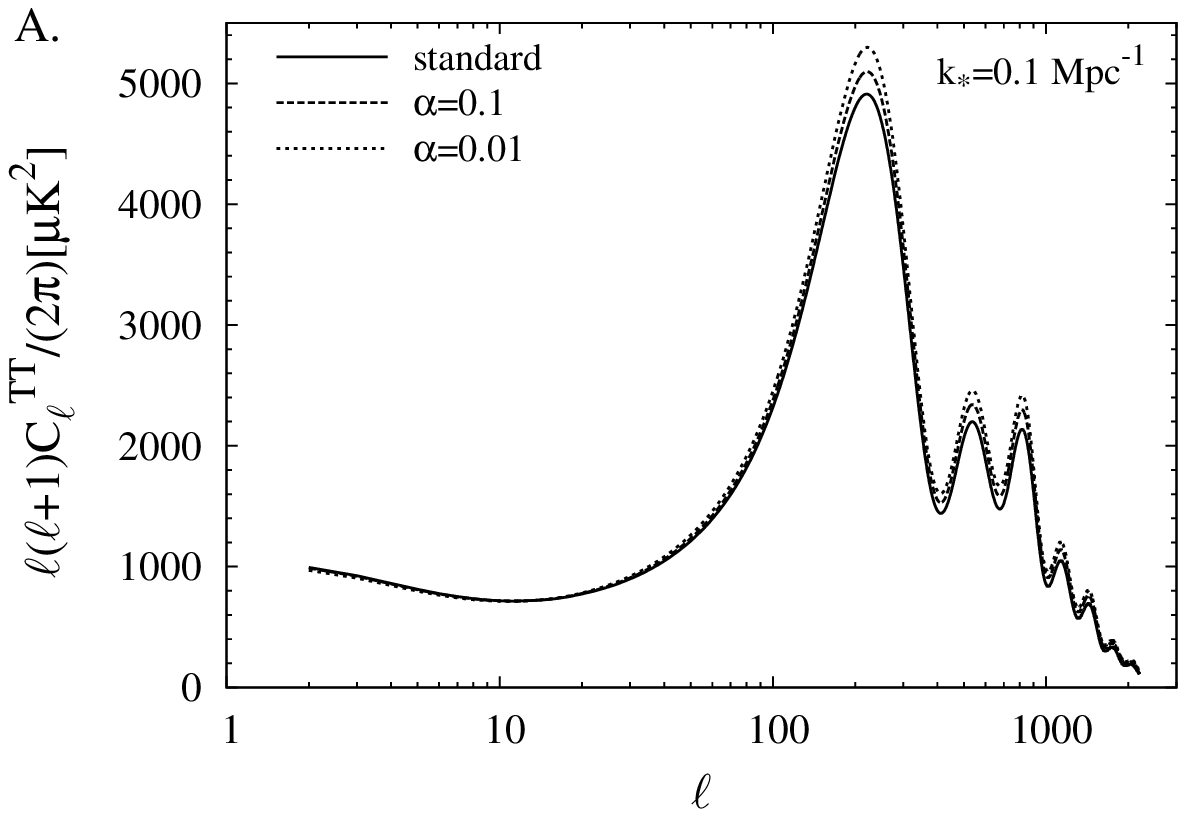}\includegraphics[width=7.7cm]{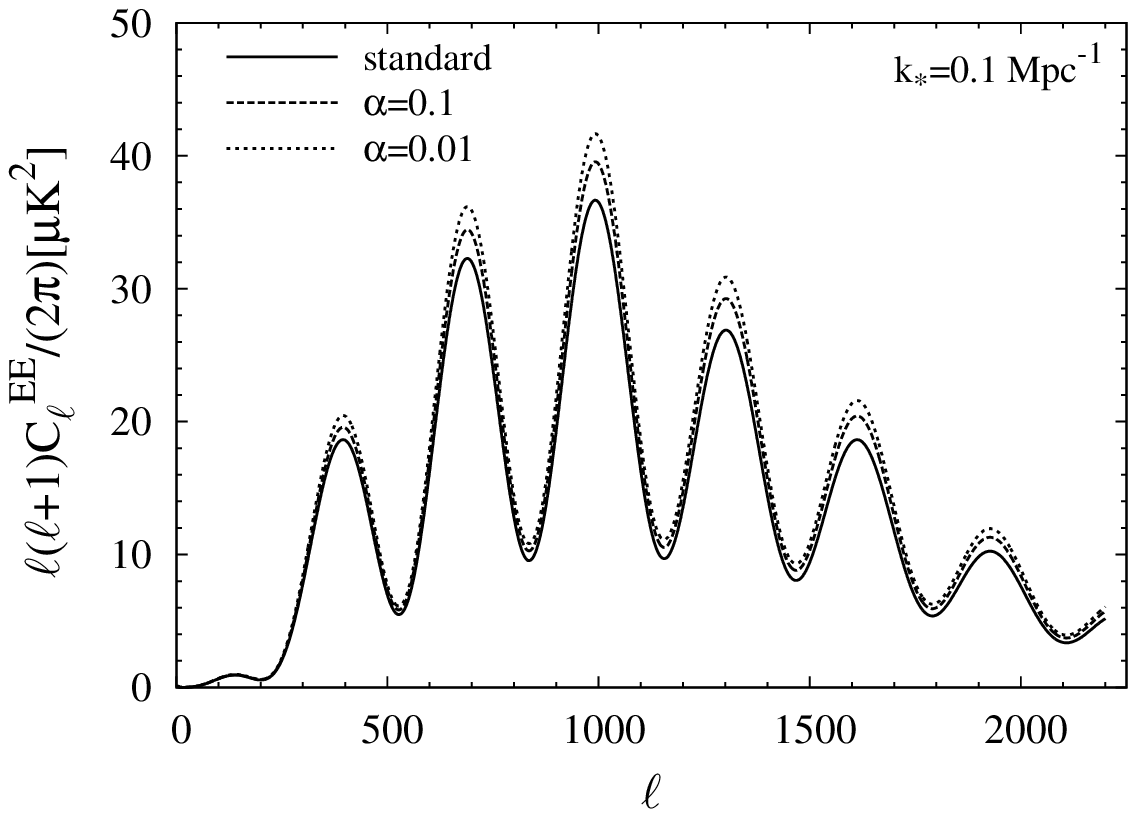}\\
\includegraphics[width=7.7cm]{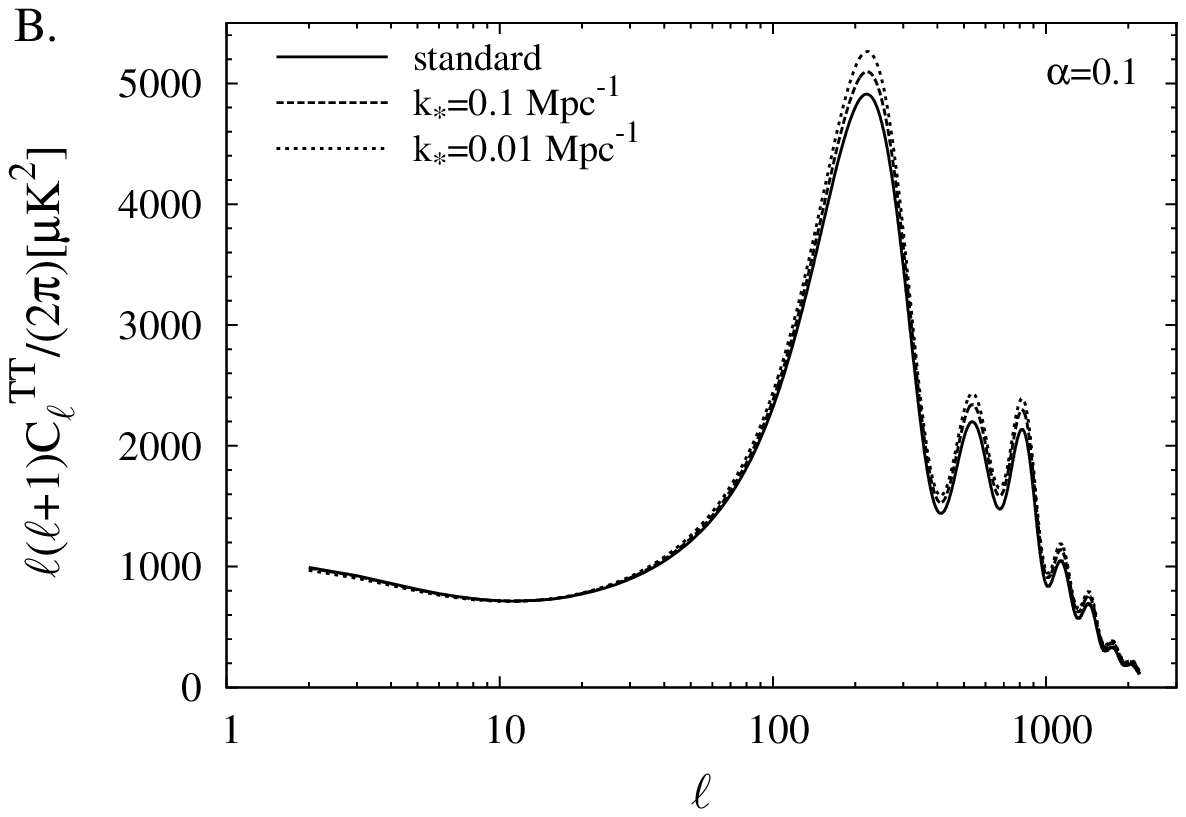}\includegraphics[width=7.7cm]{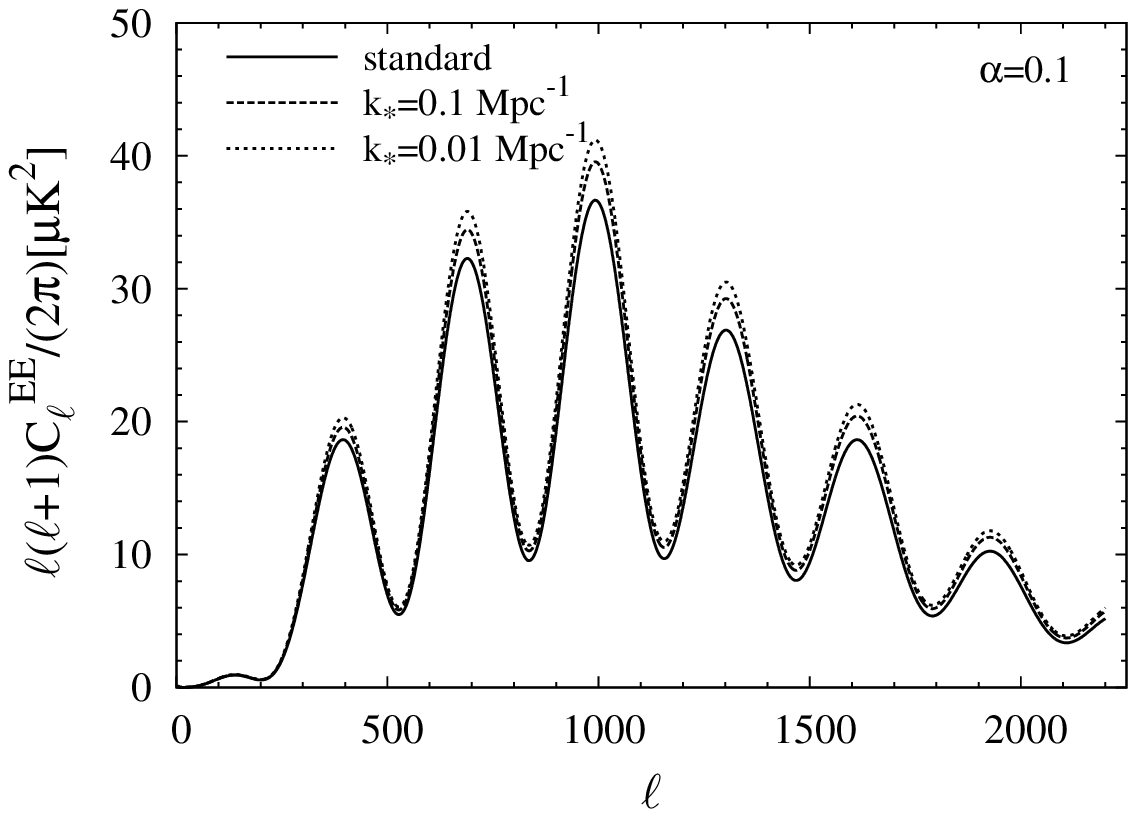}
\vspace{0.5cm}
\caption{\label{fig3} Left column: the CMB temperature power spectrum obtained by \textsc{camb} for $\cA_{\rm s}=2.1 \times 10^{-9}$, $k_0=0.002\,\text{Mpc}^{-1}$, $n_{\rm s}=0.96$ and different values of the parameters in the measure. (A) $\a=1$ (standard case, solid curve), $\a=0.1$ (dashed), $\a=0.01$ (dotted), $k_*=0.1\,\text{Mpc}^{-1}$, $A=0=B$; power increases as $\a$ decreases. (B) $\a=0.1$, $k_*=\infty$ (standard case, solid curve), $k_*=0.1\,\text{Mpc}^{-1}$ (dashed), $k_*=0.01\,\text{Mpc}^{-1}$ (dotted), $A=0=B$; multi-scale effects are triggered further to the left as $k_*$ decreases, which results in an increase of power at larger scales. Right column: the polarization $EE$ spectrum for the same choice of parameters.}
\end{figure}

\begin{figure}[ht]
\centering
\includegraphics[width=7.4cm]{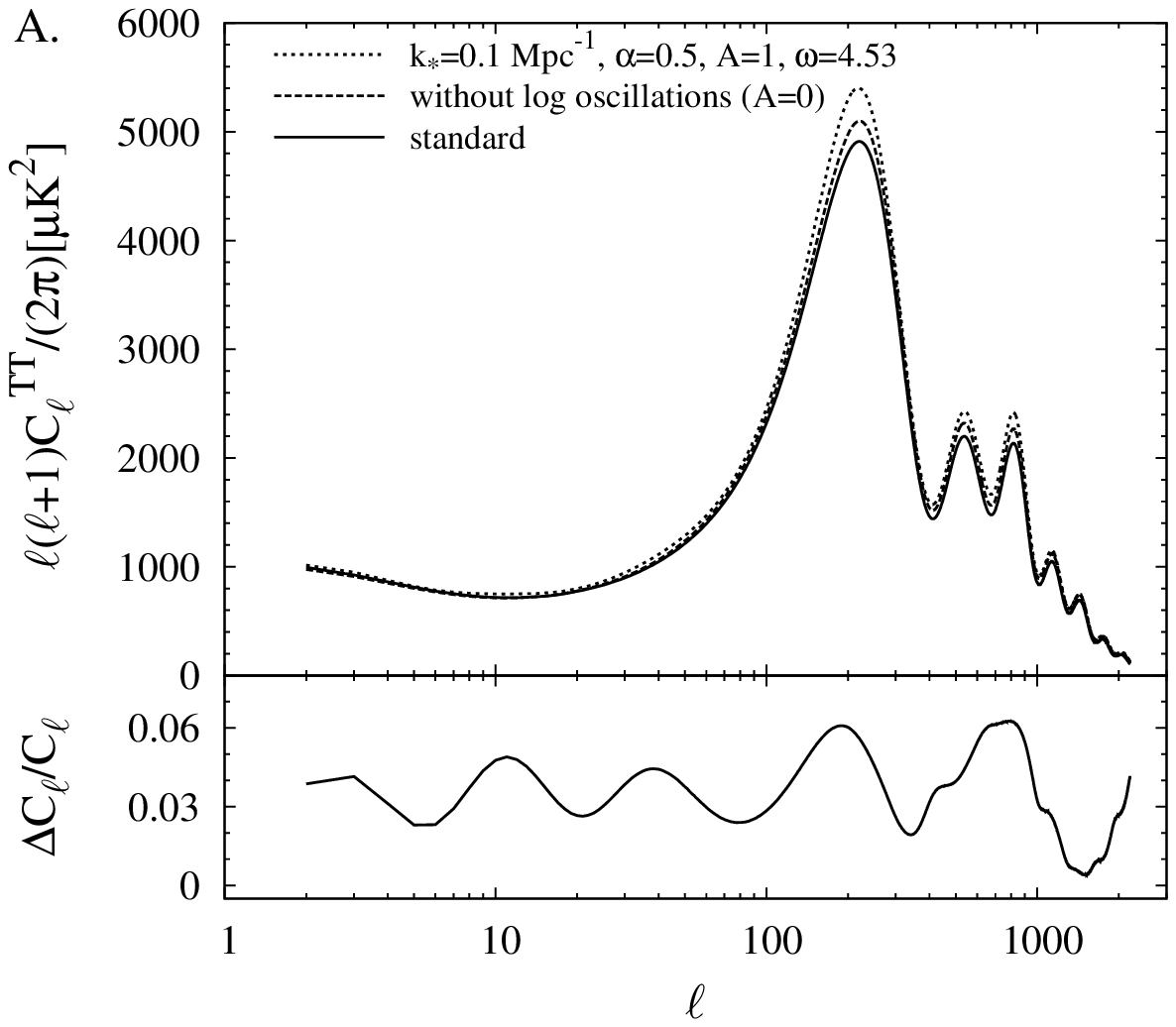}\includegraphics[width=7.4cm]{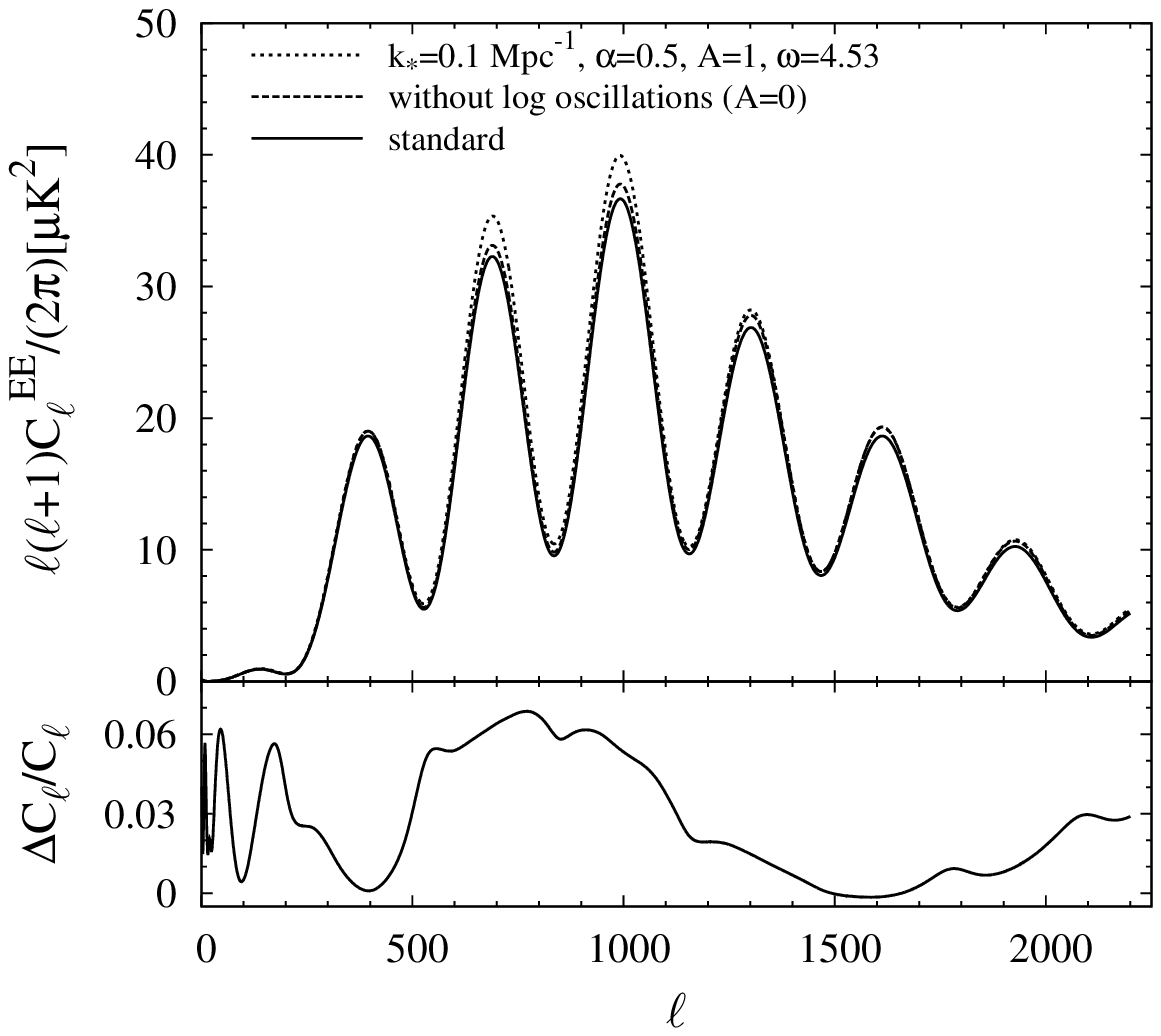}\\
\includegraphics[width=7.4cm]{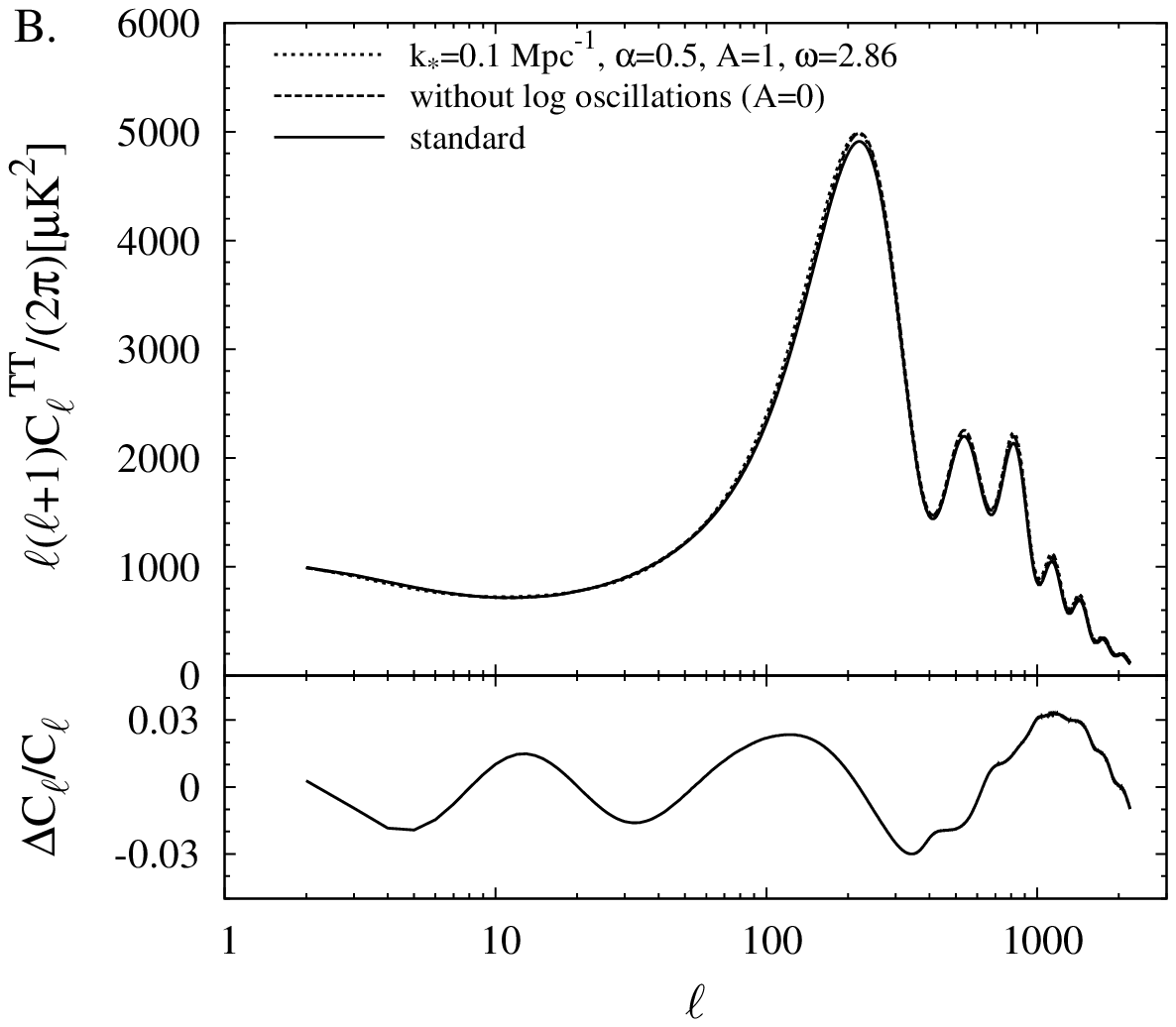}\includegraphics[width=7.4cm]{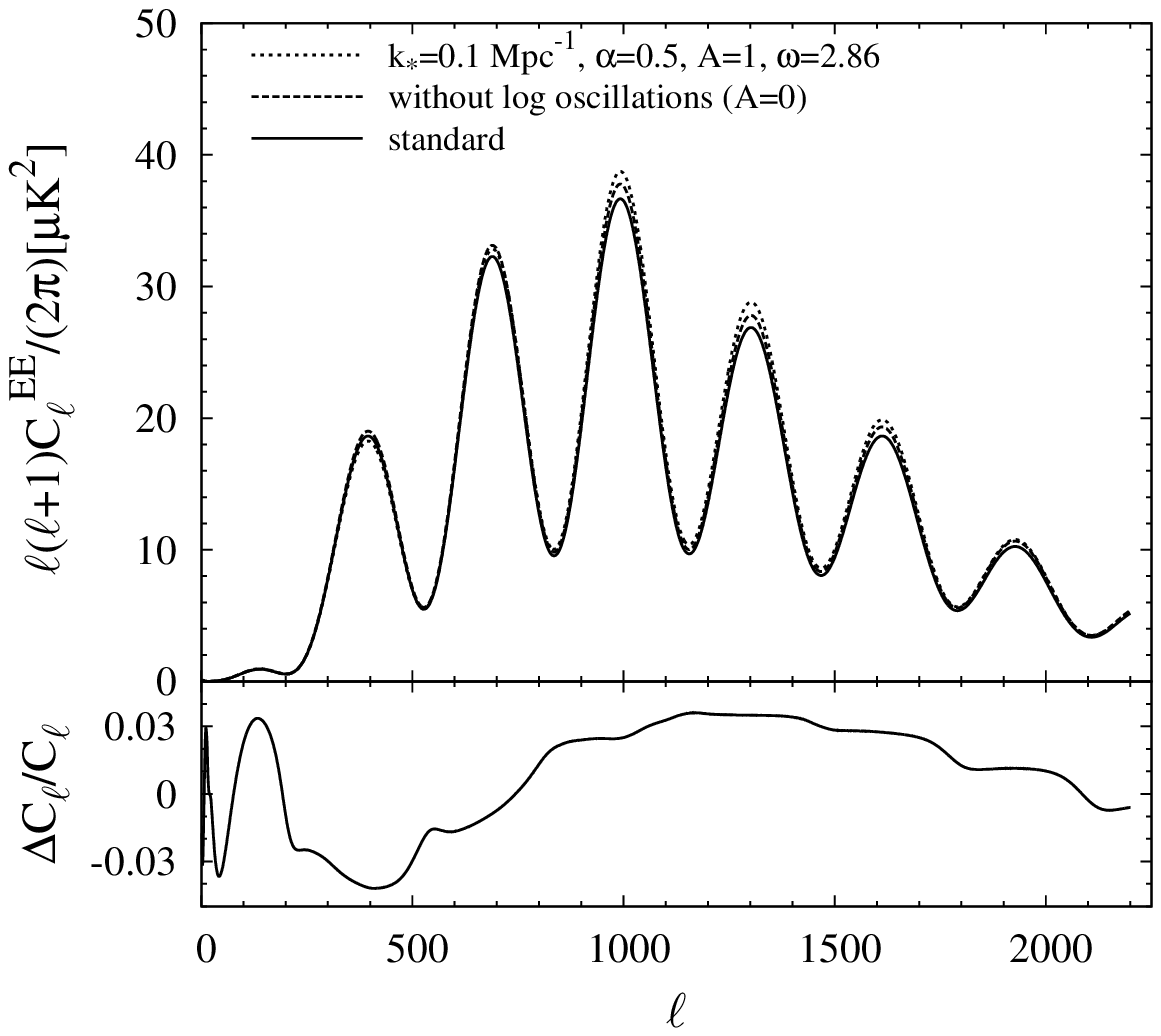}\\
\includegraphics[width=7.4cm]{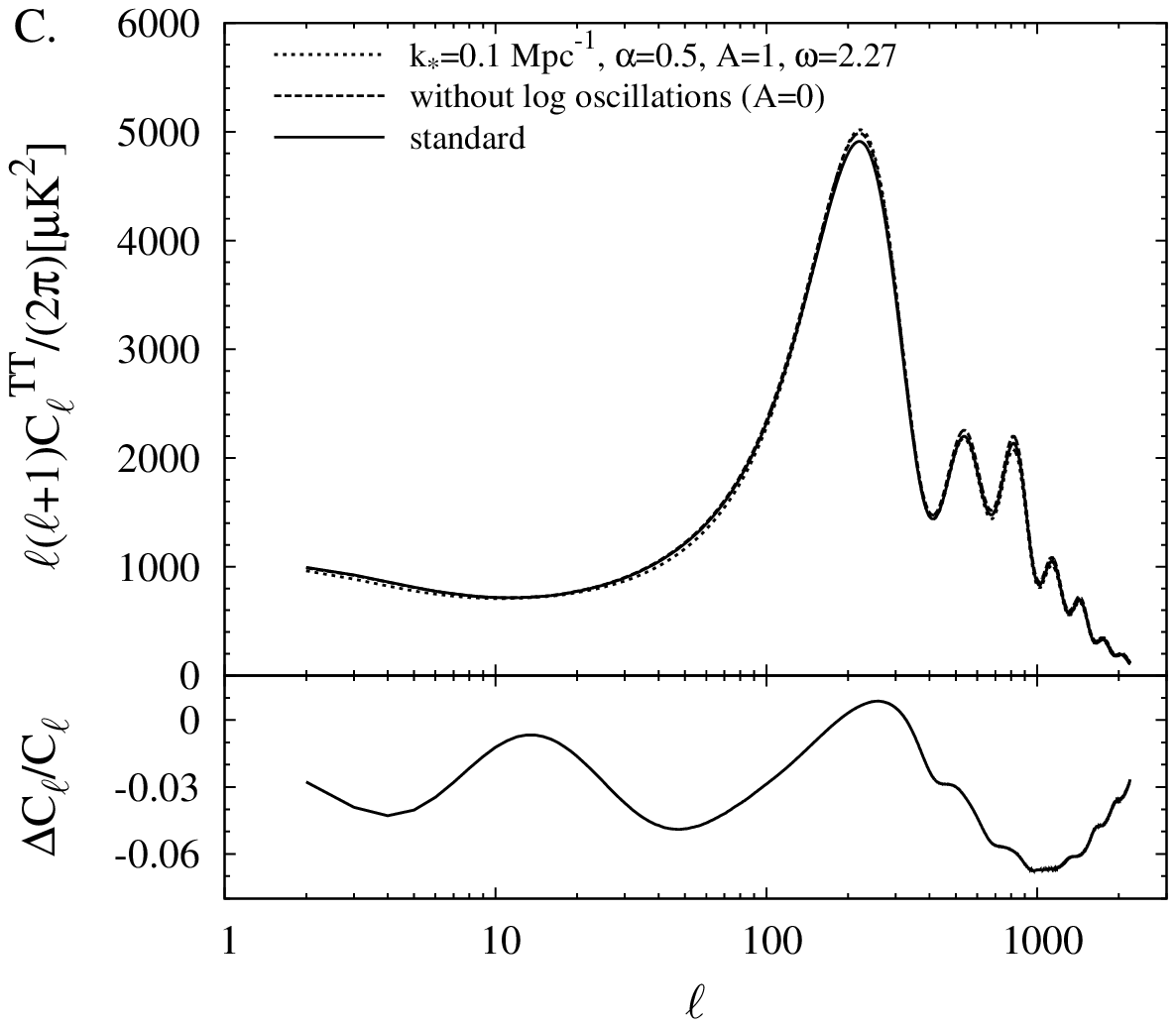}\includegraphics[width=7.4cm]{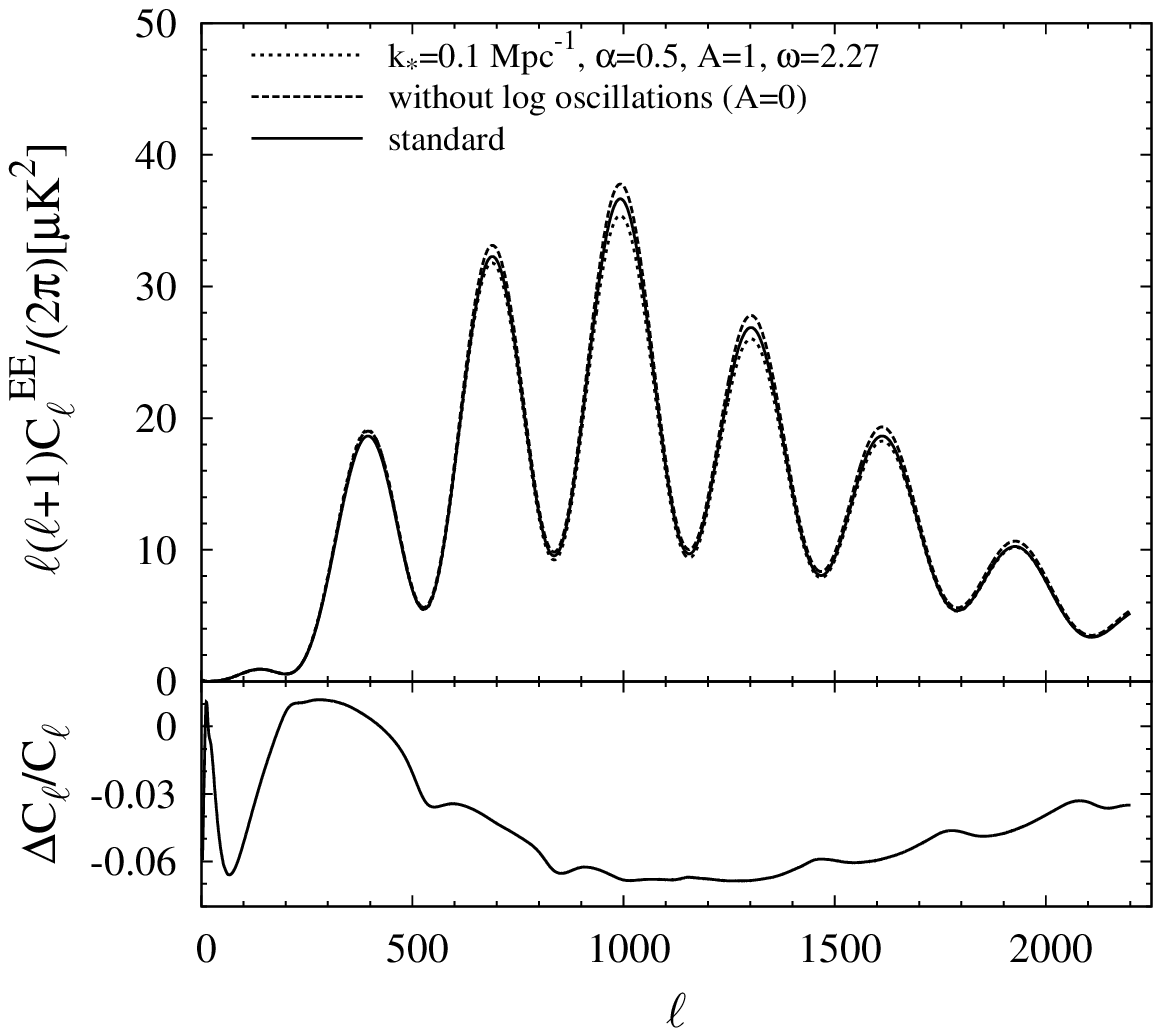}
\vspace{0.6cm}
\caption{\label{fig4} Left column: the CMB temperature power spectrum obtained by \textsc{camb} for $\cA_{\rm s}=2.1 \times 10^{-9}$, $k_0=0.002\,\text{Mpc}^{-1}$, $n_{\rm s}=0.96$, $\a=1/2$, $k_*=0.1\,\text{Mpc}^{-1}$, $B=0$, $A=0$ (dashed curve) and $A=1$ (dotted curve). (A) $N=2$, $\om=\om_2\approx4.53$. (B) $N=3$, $\om=\om_3\approx 2.86$. (C) $N=4$, $\om=\om_4\approx 2.27$; lower panels show the difference between the spectrum with log oscillations and the one without. Right column: the polarization $EE$ spectrum for the same choice of parameters.}
\end{figure}

The combined effect of the multi-fractional regimes and of the log oscillations is shown in figures \ref{fig2}--\ref{fig4}. The effect of oscillations is to modulate the zero-mode contribution; the latter gives rise to an increase in power (the greater the closer is $\a$ to zero) at small scales. A smaller $k_*$ triggers this effect at larger scales, while a large $k_*$ confines it to the rightmost part of the spectrum. The modulation of the CMB spectrum by DSI is sensitive to the relative phase between CMB peaks and the peaks of log-oscillations. When two troughs coincide, the net effect is an increase in the CMB peaks, while when a valley corresponds to the position of a CMB peak the latter undergoes a lesser increase (or even a decrease with respect to its standard value, if the amplitudes in the measure are large enough).

\subsection{Geometric power-law inflation}

In the case of multi-scale inflation, an alternative mechanism to inflation can solve the horizon problem but not the flatness problem. To see this in an alternative way than that of \cite{frc11}, let us briefly consider the model \Eq{plexp}. Expressing the temperature fluctuation spectrum $\sim [H/(2\pi v)]^2$ in terms of eq.\ \Eq{hup}, one can show that the exact scalar and tensor spectra for power-law inflation read
\ba
\cP_{\rm s}&=& \frac{\k^2}{2}(p-1)^2p\left[2^{\nu_p}\frac{\Gamma\left(\frac32+\nu_p\right)}{\Gamma\left(\frac32\right)}\right]^2\frac{1}{(2\pi q)^2}\,,\\
\cP_{\rm t}&=& 8\k^2 (p-1)^2\left[2^{\nu_p}\frac{\Gamma\left(\frac32+\nu_p\right)}{\Gamma\left(\frac32\right)}\right]^2\frac{1}{(2\pi q)^2}\,,
\ea
where
\be\label{nuep}
\nu_p=\frac{1}{p-1}\,.
\ee
In the limit $p\to\infty$, one recovers the de Sitter spectra. The tensor-to-scalar ratio is
\be
r:=\frac{\cP_{\rm t}}{\cP_{\rm s}}=\frac{16}{p}\,.
\ee
The joint \textsc{Planck}/\textsc{Bicep2}/Keck data analysis gave the upper bound $r<0.12$ at 95\% confidence level (CL) \cite{Bic15}, which translates into a lower bound for $p$:
\be
p\gtrsim 133\,.
\ee
Therefore, $p$ cannot be $O(1)$ if we want to realize viable inflation with this model, and we do need acceleration-driving matter to explain observations. Still, scale invariance can be achieved for values of $p$ smaller than in standard cosmology, where power-law inflation is actually ruled out. Another interesting feature is a scale dependence in the spectral indices, as we will see now.

From the Mukhanov--Sasaki equation \Eq{mukeq}, the horizon-crossing condition reads
\be\label{kor}
\tilde k=\frac{p-1}{p}\frac{a H}{v}\,.
\ee
We notice that
\be
\dot{\tilde k}=\tilde k H\left(1-\frac1p-\frac{\dot v}{vH}\right),
\ee
where the left-hand side is $\dot{\tilde k}=(\p_k\tilde k)\dot k$. Since
\ba
\frac{\rmd \ln\tilde k}{\rmd \ln k} &=&\tilde k\left[\frac{1}{k}+\frac{1}{k_*}\left|\frac{k_*}{k}\right|^\a F_\om(-\ln|k|)-\frac{1}{\a}\frac{k}{k_*}\left|\frac{k_*}{k}\right|^\a \p_{|k|}F_\om(-\ln|k|)\right]\nonumber\\
&=& \frac{\tilde k}{k}+\frac{\tilde k}{k_*}\left|\frac{k_*}{k}\right|^\a \left[1+A_-\cos\left(\om\ln\frac{k_\infty}{k}\right)+B_+\sin\left(\om\ln\frac{k_\infty}{k}\right)\right]\nonumber\\
&=:& f(k)\,,
\ea
we obtain
\be\label{dkt}
\frac{\rmd}{\rmd\ln k}=\frac{f(k)}{H\left(1-\frac1p-\frac{\dot v}{vH}\right)}\frac{\rmd}{\rmd t}\,,
\ee
where in the right-hand side $k=k(t)$ is a root of eq.\ \Eq{kor}. Since $\dot \cP_{\rm s,t}=-2\dot q/q=-2H/p$, we finally get the scalar and tensor spectral indices:
\ba
n_{\rm s}-1 &:=& \frac{\rmd\ln\cP_{\rm s}}{\rmd\ln k}\nonumber\\
            &=&-2\frac{f(k)}{p-1-p\dot v/(vH)}\\
&=& n_{\rm t}   :=  \frac{\rmd\ln\cP_{\rm t}}{\rmd\ln k}\,.\nonumber
\ea
In the limit $v\to 1$, the spectral indices recover their standard value $-2/(p-1)$. In the presence of anomalous-geometry effects, they acquire a highly non-trivial $k$-dependence, which can be studied more conveniently with the parametrization \Eq{Pmusc}. In that case, in fact, the main effect of the running at scales $k\gg k_*$ is summarized by the effective index \Eq{tilden}.


\section{Observational constraints from primordial spectra}\label{num}

Let us consider scalar perturbations. Equation \Eq{Pmusc} has nine free parameters: the amplitude $\cA_{\rm s}$, the spectral index $n_{\rm s}-1$, the pivot scale $k_0$ and six theoretical parameters typical of multi-scale spacetimes. The pivot scale $k_0$ is set by the experiment; in general, \textsc{Planck} results are quoted for the two values $k_{0.05}=0.05\,\text{Mpc}^{-1}$ and $k_{0.002}=0.002\,\text{Mpc}^{-1}$. Since the power spectrum at small scales ($k>k_*$) is modified by the parameter $\alpha$, we adopt the second value $k_0=k_{0.002}$. Also, the tensor-to-scalar ratio is set to zero in the analysis, since its prediction does not deviate from standard inflation.

The six multi-scale parameters are the fractional index $\a$, the scales $k_*$ and $k_\infty$, the amplitudes $A$ and $B$ and the frequency $\om$. In the absence of oscillations, there are only two free parameters but some theoretical considerations suggest $\a=1/2$ as a special value which drives the dimensionality of spacetime to 2 in the ultraviolet \cite{frc1,frc2,frc7}. The simulations we have run are: {\bf (i)} $k_*$ free, $\a=1/2$, $A=0=B$; {\bf (ii)} $k_*$ and $\a$ free, $A=0=B$.

When the amplitudes are non-zero, in order to make the likelihood analysis manageable we can fix two of the parameters. To begin with, we take the microscopic scale $E_\infty$ at the bottom of the multi-fractal hierarchy to be the Planck energy \cite{ACOS}, in which case
\be
k_\infty =\frac{\sqrt{3}}{\lp}\approx 3.3\times 10^{60}~\text{Mpc}^{-1}\,.
\ee
Second, we performed a likelihood analysis leaving all the other parameters free with theoretically motivated priors
\be
k_*>0\,,\qquad 0<\a,A,B<1\,,\qquad 0<\om<\om_2\,. 
\ee
For simplicity, we varied $\om$ on the continuous prior $0<\om<\om_2$ rather than on the discrete spectrum \Eq{omspe}, where $\om_2\approx 4.53$ is the highest value in the spectrum \Eq{omspe2}. We found strong oscillations in the distribution of $\om$, while the likelihood distributions of the other parameters of the measure did not peak at any specific value. This strong modulation of the likelihood distribution of some parameters is a known feature of log-oscillating spectra \cite{MaRi1,MaRi2,EKP}. However, in our case the peaks in the distribution of $\om$ do not correspond to any of the theoretically allowed discrete values and, moreover, by changing the allowed range of $A$ and $B$ one can check that the $\om$ distribution strongly depends on the priors. Therefore, we conducted separate simulations for fixed values of $\om$: {\bf (iii)}--{\bf (v)} $N=2,3,4$ in equation \Eq{omspe}, $k_*$, $\a$, $A$ and $B$ free. Due to the high degeneracy of viable spectra, at first it is not advisable to fix the parameters further (for instance, by taking $\a=1/2$ or one of the amplitudes equal to zero). Also, a very large $N$ would correspond to low-frequency oscillations and it would just affect the spectral amplitude; for this reason, we explore only the first three values of the frequency $\om_N$. However, we will fix $\a$ in some controllable cases with {\bf (vi)}--{\bf (viii)} $\a=1/2$ and $N=2,3,4$ and {\bf (ix)}--{\bf (xi)} $\a=0.1$ and $N=2,3,4$. The advantage in doing so is that, after checking consistency with the non-marginalized results, it is possible to extract stringent bounds (the first of this kind for this theory) on the amplitudes $A$ and $B$ of the log-oscillating measure.

We checked that inclusion of polarization data improves the marginalized likelihood contours, while lensing data worsen them making them patchy and less sharp. In the following, we use temperature and polarization data (the ``TT+lowP'' dataset as dubbed by the \textsc{Planck} team).


\subsection{Without oscillations}

\begin{figure}[ht]
\centering
\includegraphics[width=8.5cm]{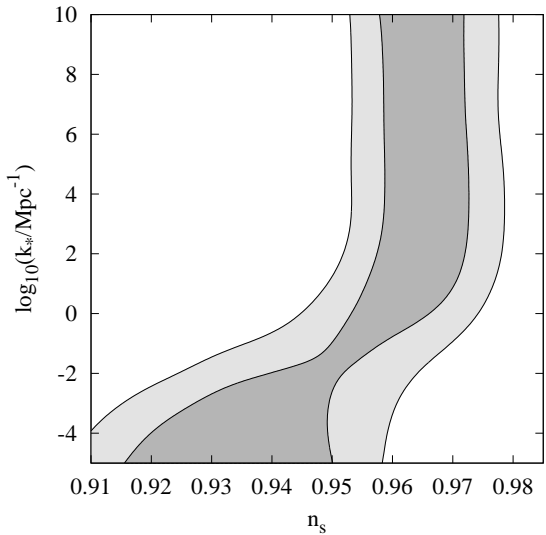}
\caption{\label{fig5} Likelihood contours in the $(n_{\rm s},k_*)$ plane for $k_0=0.002\,\text{Mpc}^{-1}$, $\a=1/2$, $A=0=B$ and a tensor-to-scalar ratio $r=0$.}
\end{figure}

Case {\bf (i)} is preliminary to all the others and explores the degeneracy in the parameter space induced by the scale $k_*$. For $\a=1/2$, the likelihood contour bounds obtained with the \textsc{CosmoMC} code are shown in figure \ref{fig5}.
The corridor of allowed values of the characteristic scale extends from arbitrarily small to arbitrarily large $k_*$. The spectrum \Eq{Pmusc} and the form \Eq{tilden} of the effective spectral index clarify why. For large $k_*$,
\be\label{staspe}
\cP_{\rm s}\simeq \cA_{\rm s}\left(\frac{k}{k_0}\right)^{n_{\rm s}-1}.
\ee
and one recovers the usual spectrum with $n_{\rm s}\approx 0.97$--$0.98$. For small $k_*$ ($\ln k_*\to - \infty$ in the figure), 
\be\label{nstaspe}
\cP_{\rm s}\simeq \cA_{\rm s}\left(\frac{k}{k_0}\right)^{\a(n_{\rm s}-1)}
\ee
and the spectral index $n_{\rm s}$ can tolerate large deviations from unity. Thus, we can achieve viable inflation even away from slow roll. Note that we have set $r=0$ and ignored, due to the lack of any strong constraint on the tensor index, the information from the inflationary consistency relation $r=-8 n_{\rm t}$.

\begin{figure}[ht]
\centering
\includegraphics[width=7.6cm]{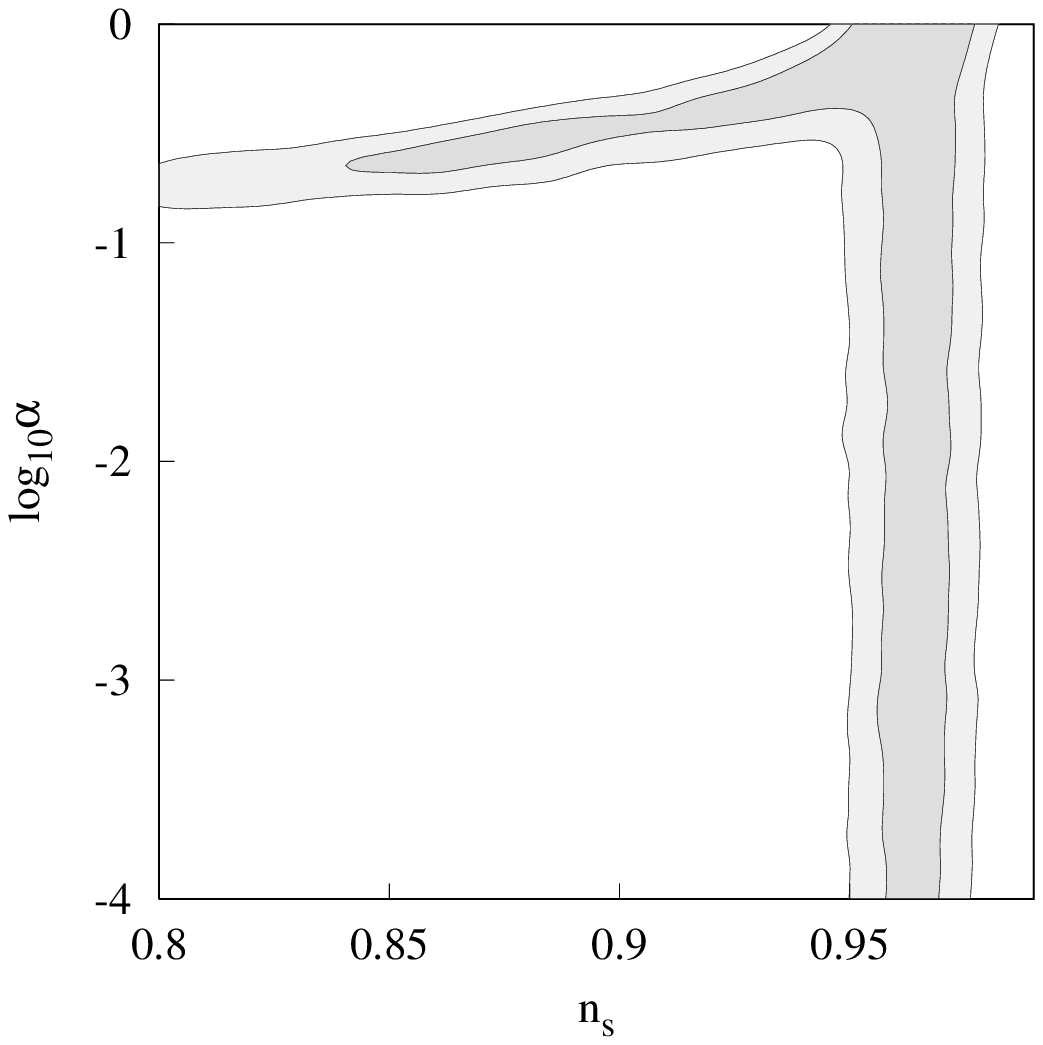}
\includegraphics[width=7.6cm]{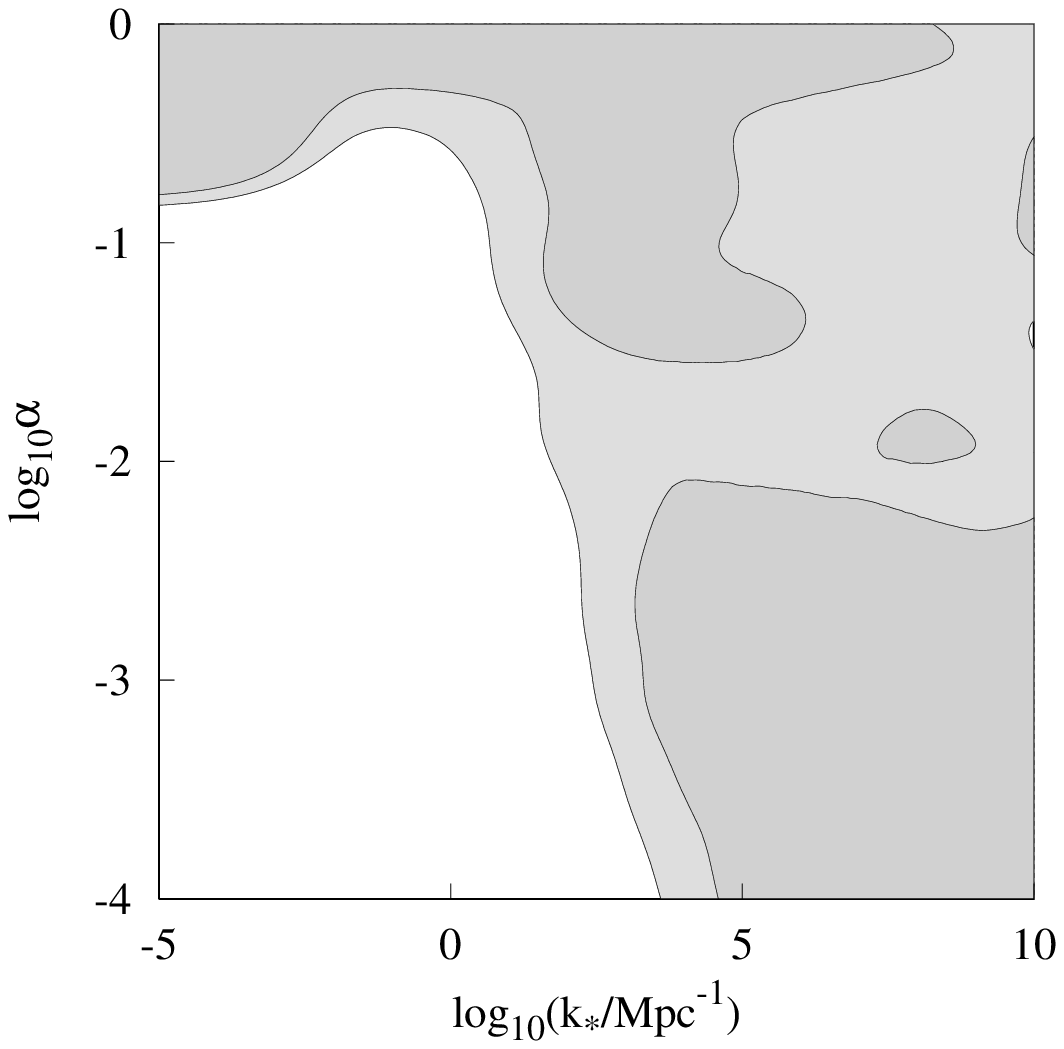}
\caption{\label{fig6} Likelihood contours in the $(n_{\rm s},\a)$ plane (left panel) and in the $(k_*,\a)$ plane (right panel), with $k_0=0.002\,\text{Mpc}^{-1}$, $A=0=B$ and a tensor-to-scalar ratio $r=0$.}
\end{figure}

\begin{figure}[ht]
\centering
\includegraphics[width=17cm]{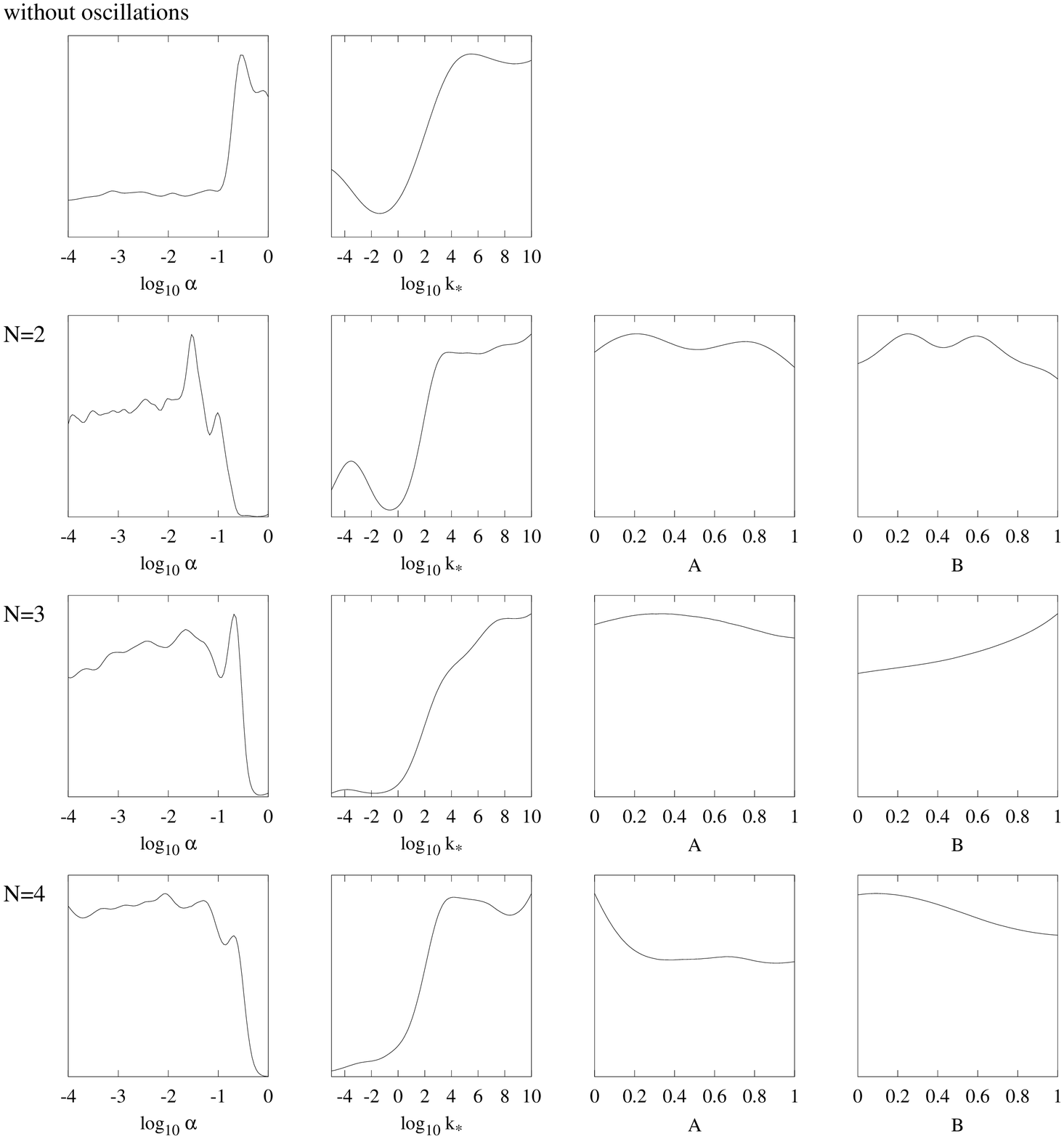}
\caption{\label{fig7} One-dimensional likelihood contours for some of the cosmological parameters, with $k_0=0.002\,\text{Mpc}^{-1}$ and a tensor-to-scalar ratio $r=0$.
Row 1: contours for $\a$ and $k_*$ without oscillations ($A=0=B$). Rows 2--4: contours for $\a$, $k_*$, $A$ and $B$ with $N=2,3,4$ in equation \Eq{omspe}.}
\end{figure}

The form \Eq{staspe}--\Eq{nstaspe} of the spectrum is a so-called broken power law. Since it is already known that phenomenological broken power laws can fit well \textsc{Planck} data \cite{PXX}, our concern here is about the details of the viable parameter space. To check the robustness of our results against marginalization artifacts, we move to case {\bf (ii)} where also $\a$ varies (figures \ref{fig6} and \ref{fig7}).

In the left panel of figure \ref{fig6}, we have marginalized over $k_*$. Near $n_{\rm s}\approx 0.96$ (vertical corridor, large $k_*$), we have the standard result: there is a degeneracy along the $\a$-direction because the primordial index $n_{\rm s}-1$ becomes close to scale invariance and the effective index \Eq{tilden} remains scale invariant for any $0<\a<1$. For $-1.1<\log_{10}\a<-0.5$ (corresponding to $0.08<\a<0.32$), the fractional exponent $\a$ is sufficiently small to allow for as small an index as $n_{\rm s}\approx 0.8$. The horizontal corridor (small $k_*$) continues to the left as $\a$ further decreases to zero. The value $\a=1/2$ corresponds to $\log_{10}\a\approx -0.30$.

In the right panel of figure \ref{fig6}, we have marginalized over $n_{\rm s}$. For large $k_*$, we have the standard spectrum \Eq{staspe} insensitive to the value of $\a$, while for small $k_*$ we discover that $\a$ cannot be arbitrarily small, lest \Eq{nstaspe} get too close to the experimentally excluded Harrison--Zel'dovich spectrum. This determines a lower bound on the fractional exponent: $\log_{10}\a>-1$, i.e.,
\be\label{lowbaa}
\a>0.1\,.
\ee
At first, this result seems promising because it is in agreement with the one-dimensional contour presented in the first panel of figure \ref{fig7} which shows that the $\a$ likelihood has a sharp increase for $\log_{10}\a\gtrsim -1$. However, we will see in section \ref{disc} that the bound \Eq{lowbaa} should be discarded when independent constraints on the parameter space are taken into account.


\subsection{With oscillations}\label{withosc}

\begin{figure}[ht]
\centering
\includegraphics[width=6.9cm]{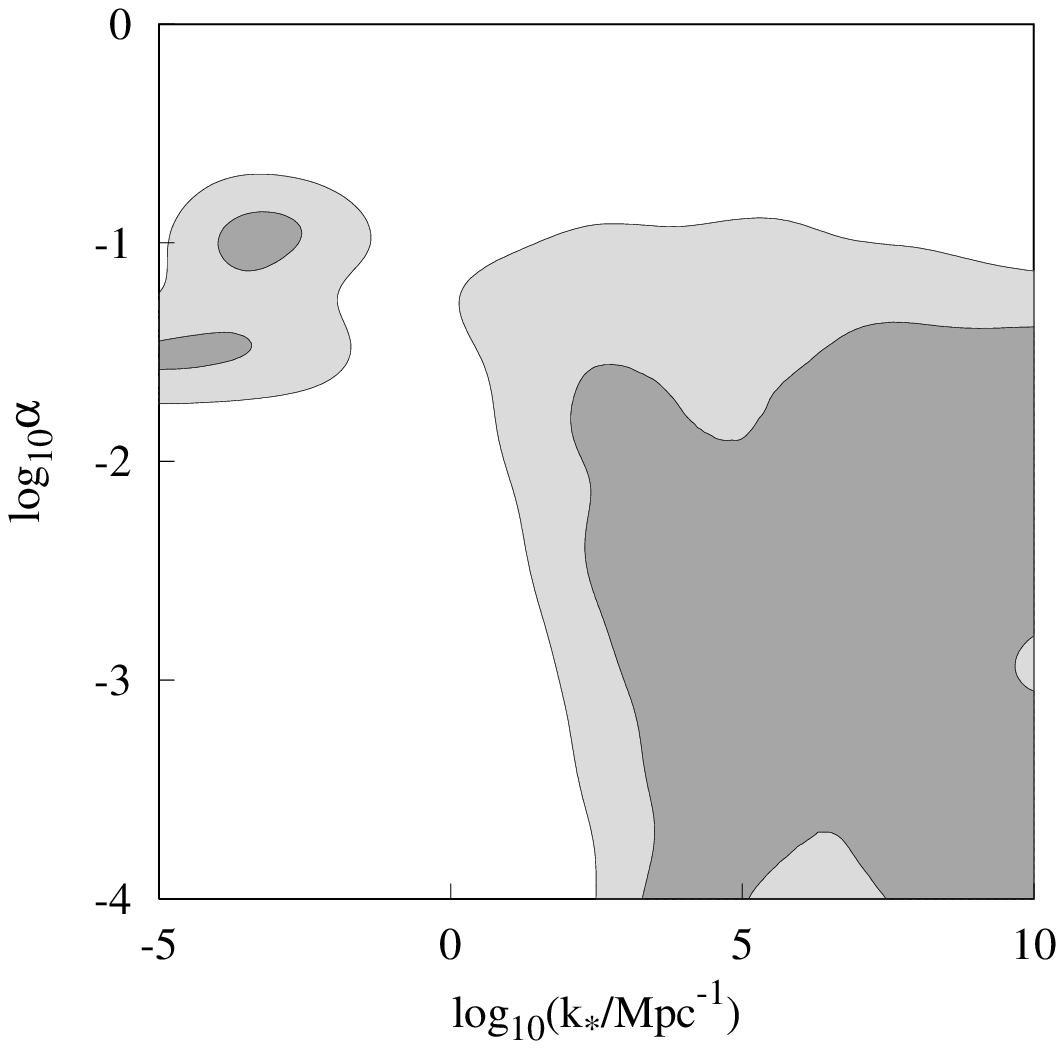}\includegraphics[width=6.9cm]{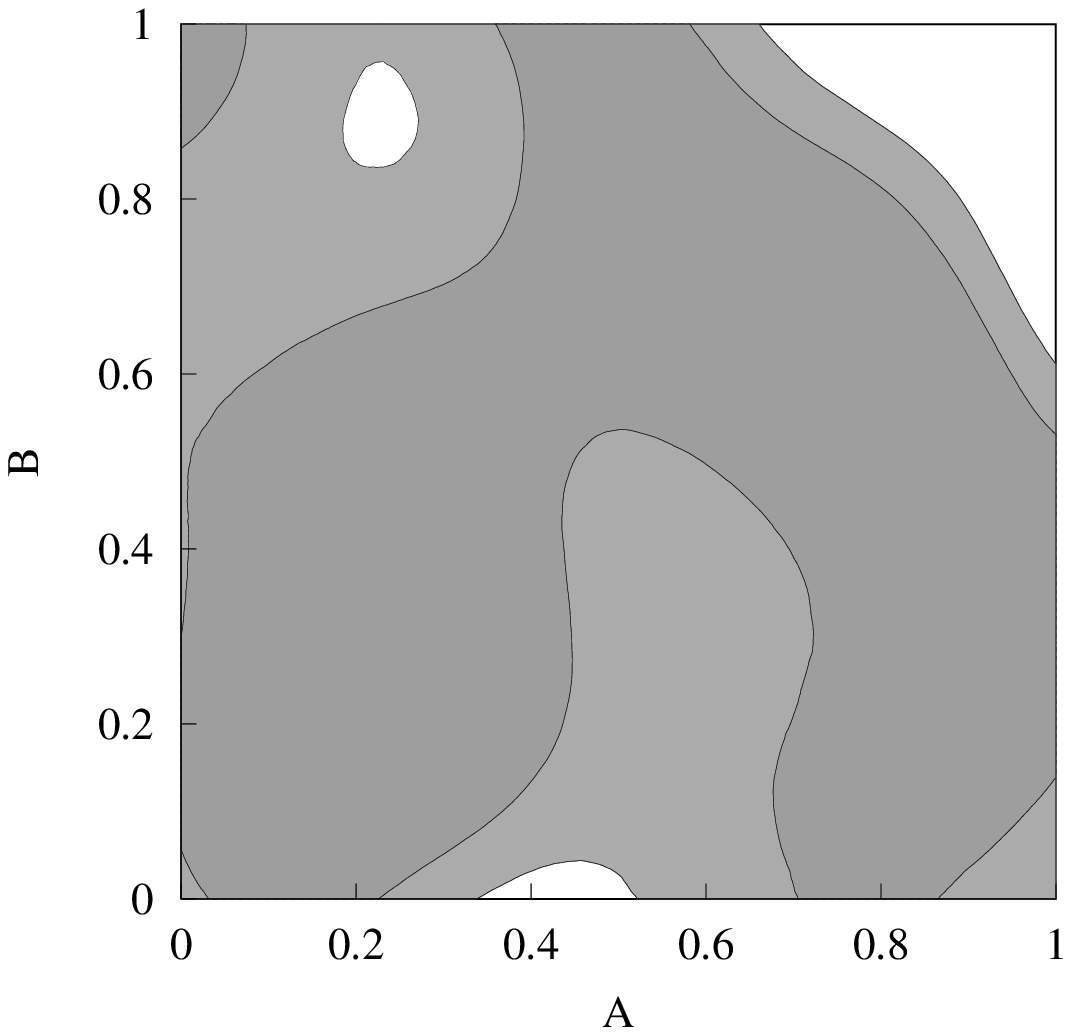}\\
\includegraphics[width=6.9cm]{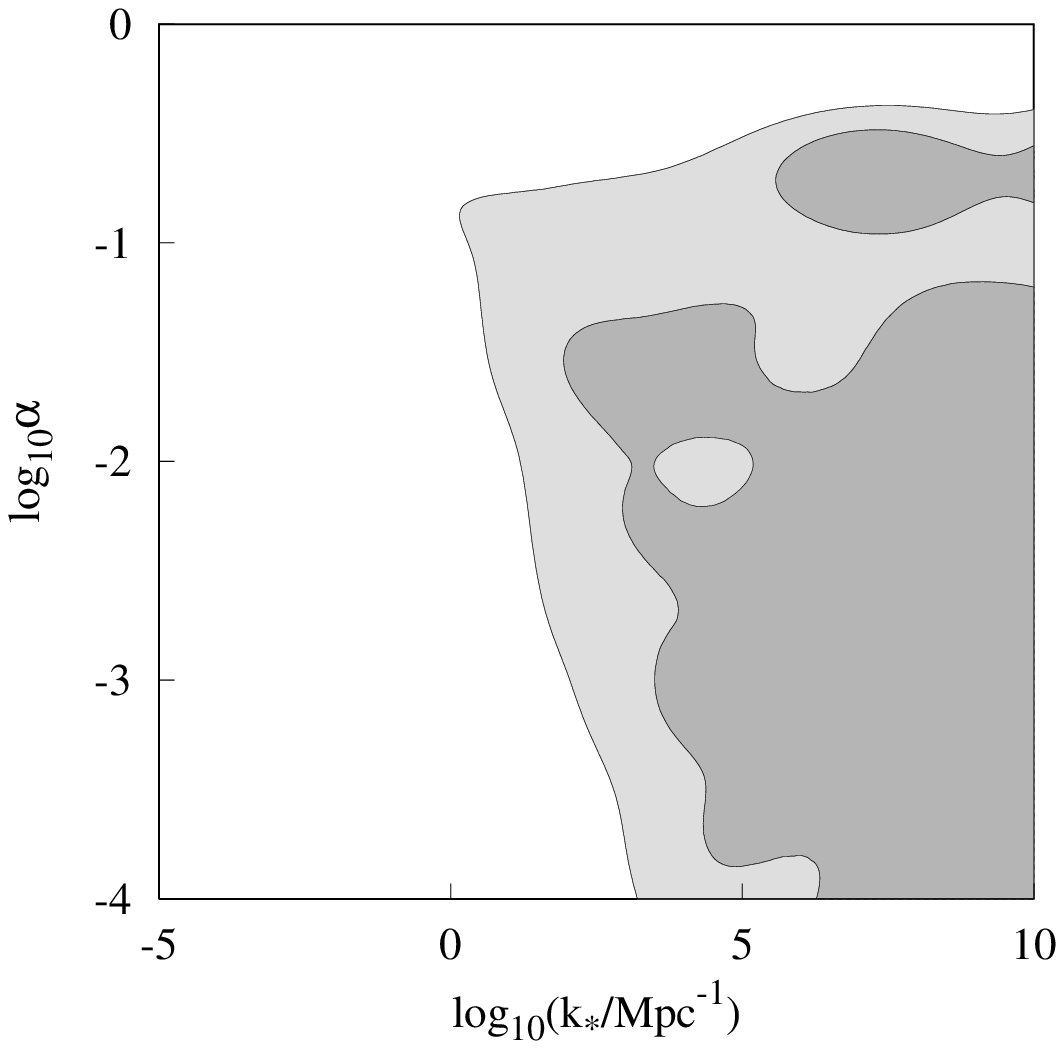}\includegraphics[width=6.9cm]{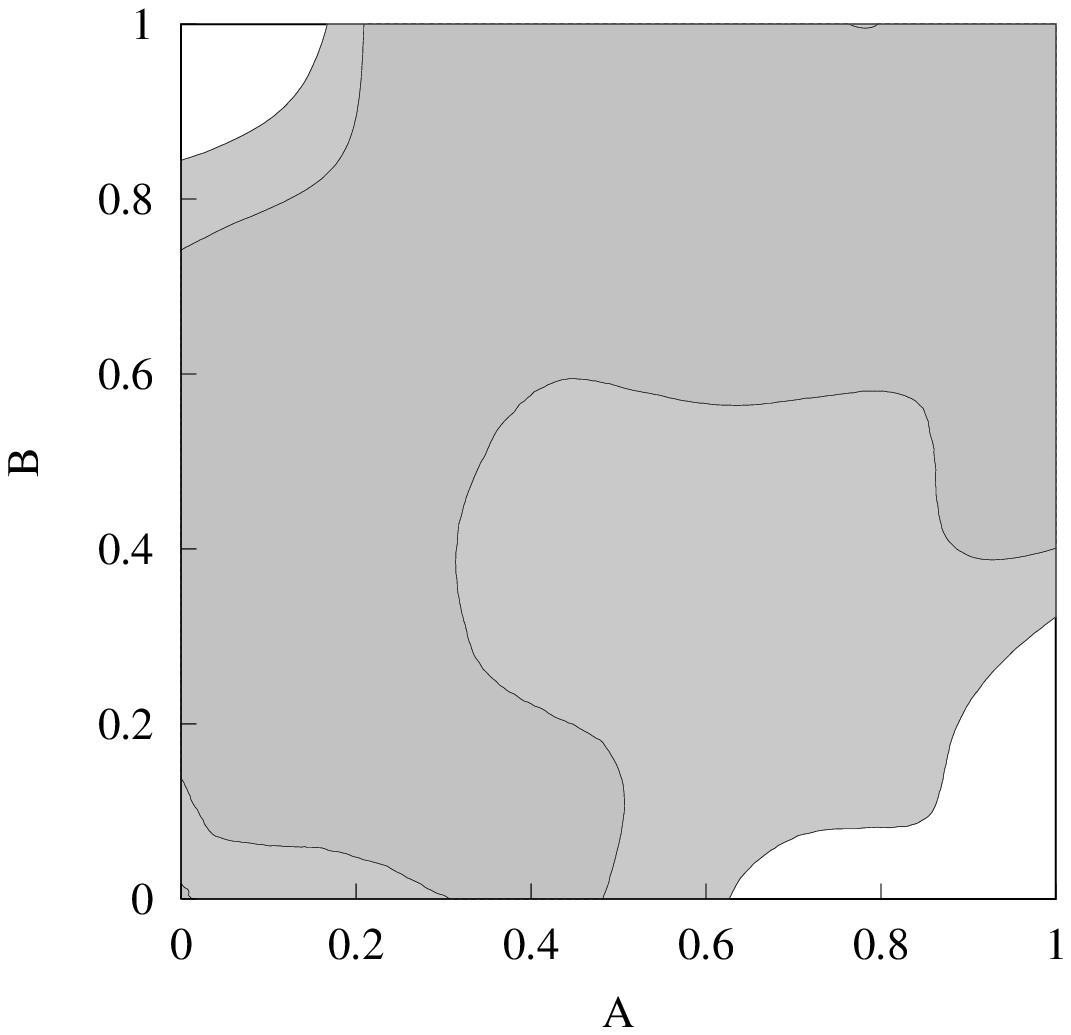}\\
\includegraphics[width=6.9cm]{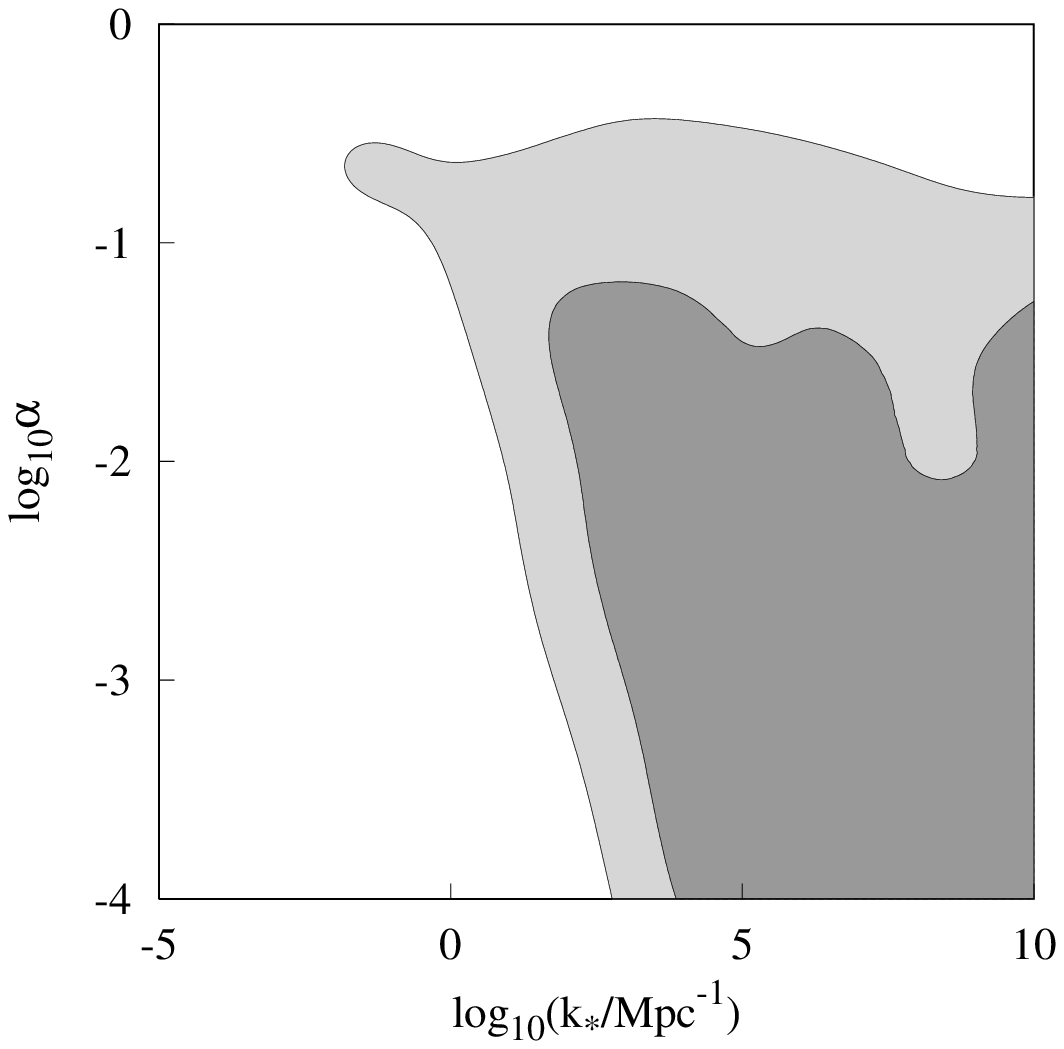}\includegraphics[width=6.9cm]{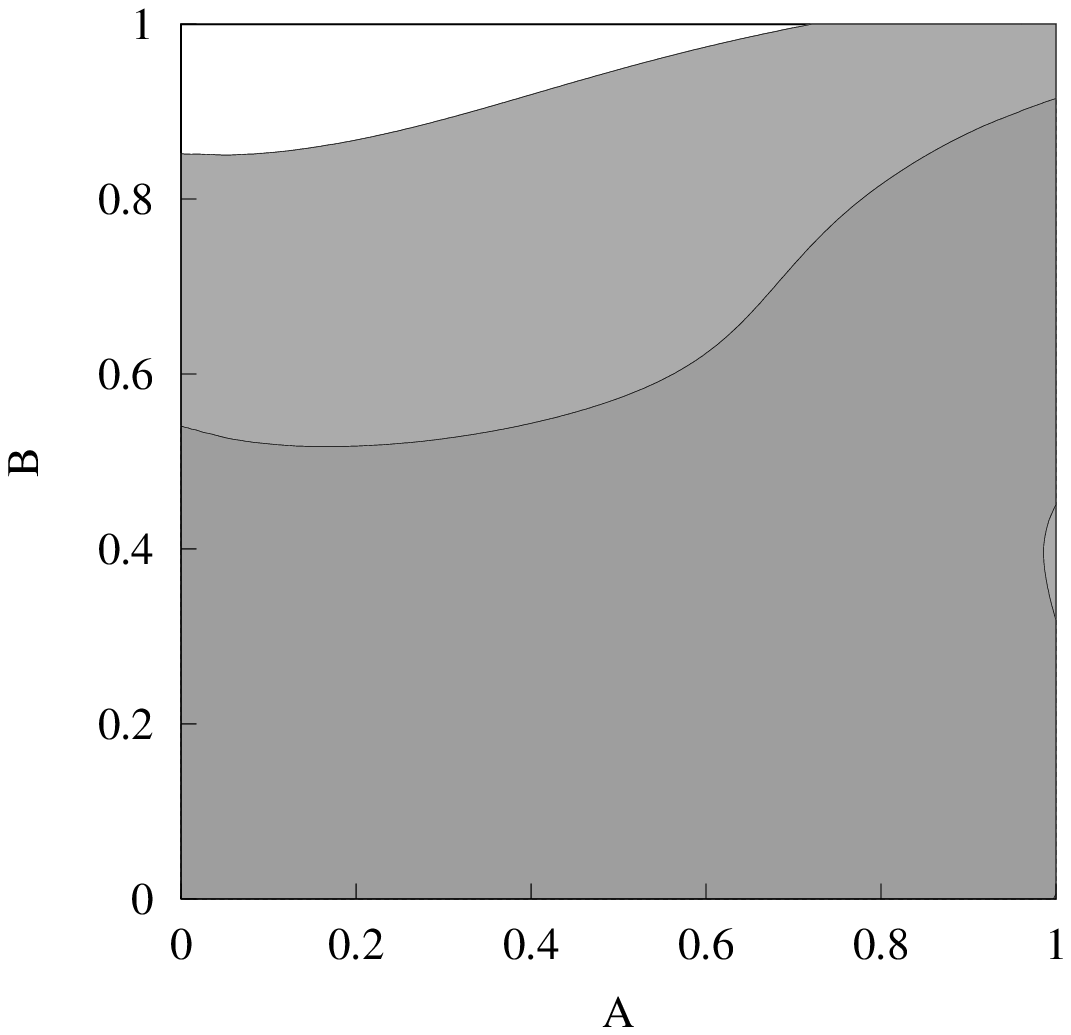}
\caption{\label{fig8} Top to bottom: marginalized likelihood in the $(k_*,\a)$ and $(A,B)$ planes for $N=2,3,4$.}
\end{figure}

In all the cases {\bf (iii)}--{\bf (v)}, we let both $A$ and $B$ free at the same time because fixing only one of the amplitudes would correspond to fix constant phases in the oscillations, a procedure that would lead to artifacts. We checked this explicitly. We do not show the likelihood profiles of the numerical runs with free $\om_N$ or fixed $A$ or $B$ since they do not give significant information. Figure \ref{fig7} collects some of the marginalized one-dimensional likelihood profiles for $N$ fixed. Interestingly, the case $\alpha=1$ is excluded by this analysis, mainly because oscillations are severely constrained by CMB data when they are the only modification to the power spectrum. The oscillation frequency $\om_N$ becomes very small as $\alpha$ decreases and the spectrum has less oscillations. The marginalized likelihood in the $(k_*,\a)$ and $(A,B)$ planes for $N=2,3,4$ is shown in figure \ref{fig8}.

Bounds on $A$ and $B$ (rightmost panels in figure \ref{fig8}) are very weak because the analysis includes both small and large values of $\om_N$. When $omega_N$ is small, the effect is similar to changing either the overall spectral amplitude or the tilt, so that different values of $A$ and $B$ produce spectra degenerate with those obtained by varying $\mathcal{A}_{\rm s}$ and $n_{\rm s}$ and cannot be constrained strongly. Nevertheless, some small portions of the square $[0,1]\times[0,1]\ni(A,B)$ are excluded at the 95\% CL.

The bounds on $\a$ and $k_*$ are sharper. For $N=2,3,4$, we find about the same upper limit on $\a$ and lower limit on $k_*$, which can be extracted from the figure. At the 95\% CL, we have $\log_{10}\a\lesssim -1,-0.2,-0.25$ for $N=2,3,4$, respectively, while in general $\log_{10}(k_*\,{\rm Mpc})\gtrsim 2$. Therefore,
\be\label{lowbaa2}
\boxd{\a\lesssim 0.1-0.6\,,\qquad  k_*>10^2\,{\rm Mpc}^{-1}\,.}
\ee

To summarize so far, we cannot constrain efficiently the model without log oscillations from the Monte Carlo analysis: the lowest allowed value for $k_*$ is typically small and, as we shall see, it corresponds to a low characteristic energy $E_*$. Moreover, and generally speaking, with respect to the range of scales of the CMB spectrum, in the case with oscillations the frequencies in the spectrum \Eq{omspe} are too low to produce an observable modulation pattern for $\a\lesssim 0.5$. This is consistent with the results of \cite{HHSW}, where it was shown that log oscillations could be detectable only if their frequency is sufficiently high, $\om\sim O(1)-O(10)$. 

\begin{figure}[ht]
\centering
\includegraphics[width=10cm]{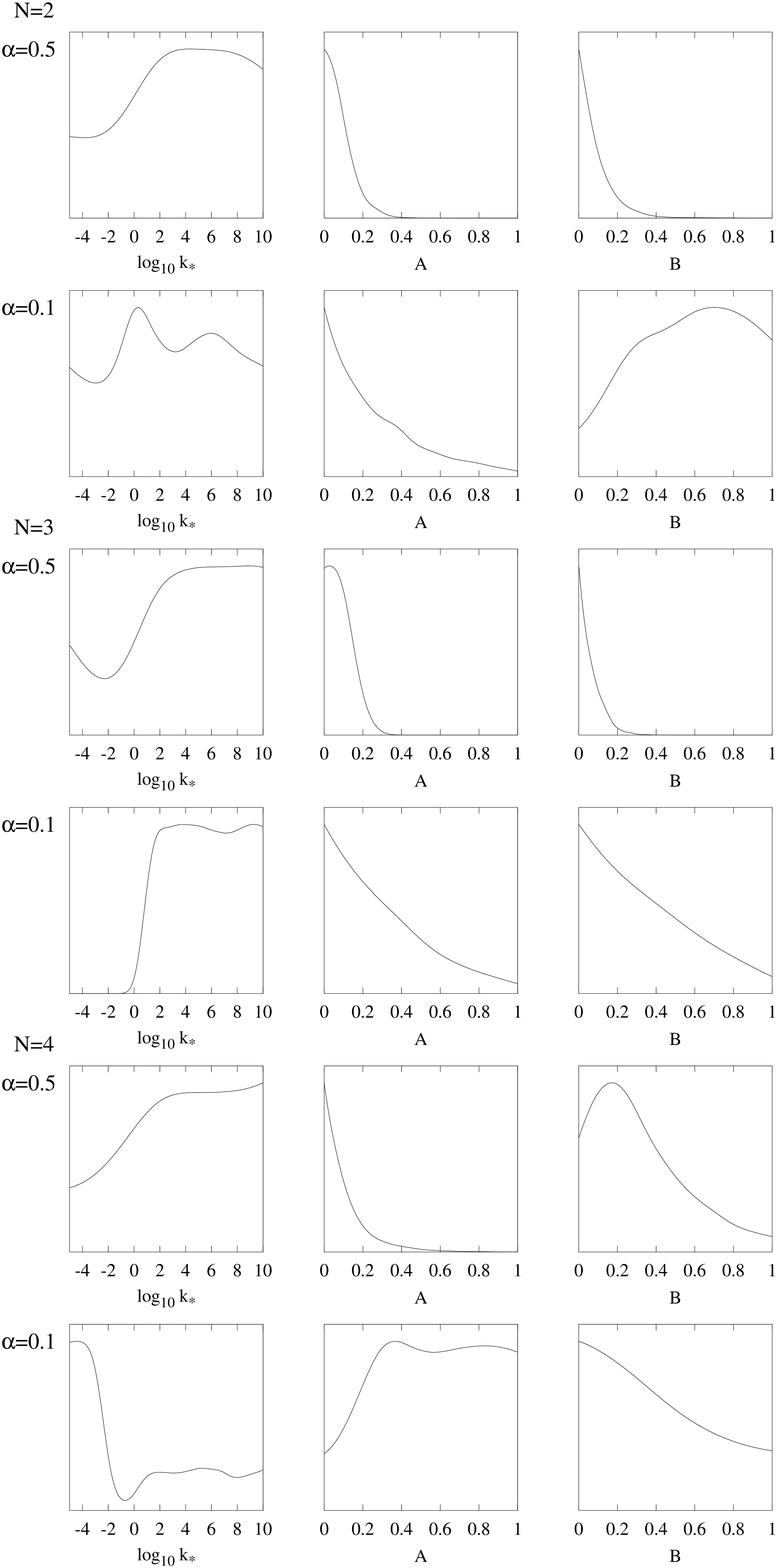}
\caption{\label{fig9} One-dimensional likelihood contours for some of the cosmological parameters, with $k_0=0.002\,\text{Mpc}^{-1}$, a tensor-to-scalar ratio $r=0$ and $\a=0.5$ or $\a=0.1$. The three six-panel blocks of contours correspond to $N=2,3,4$.}
\end{figure}

\begin{figure}[ht]
\centering
\includegraphics[width=6.8cm]{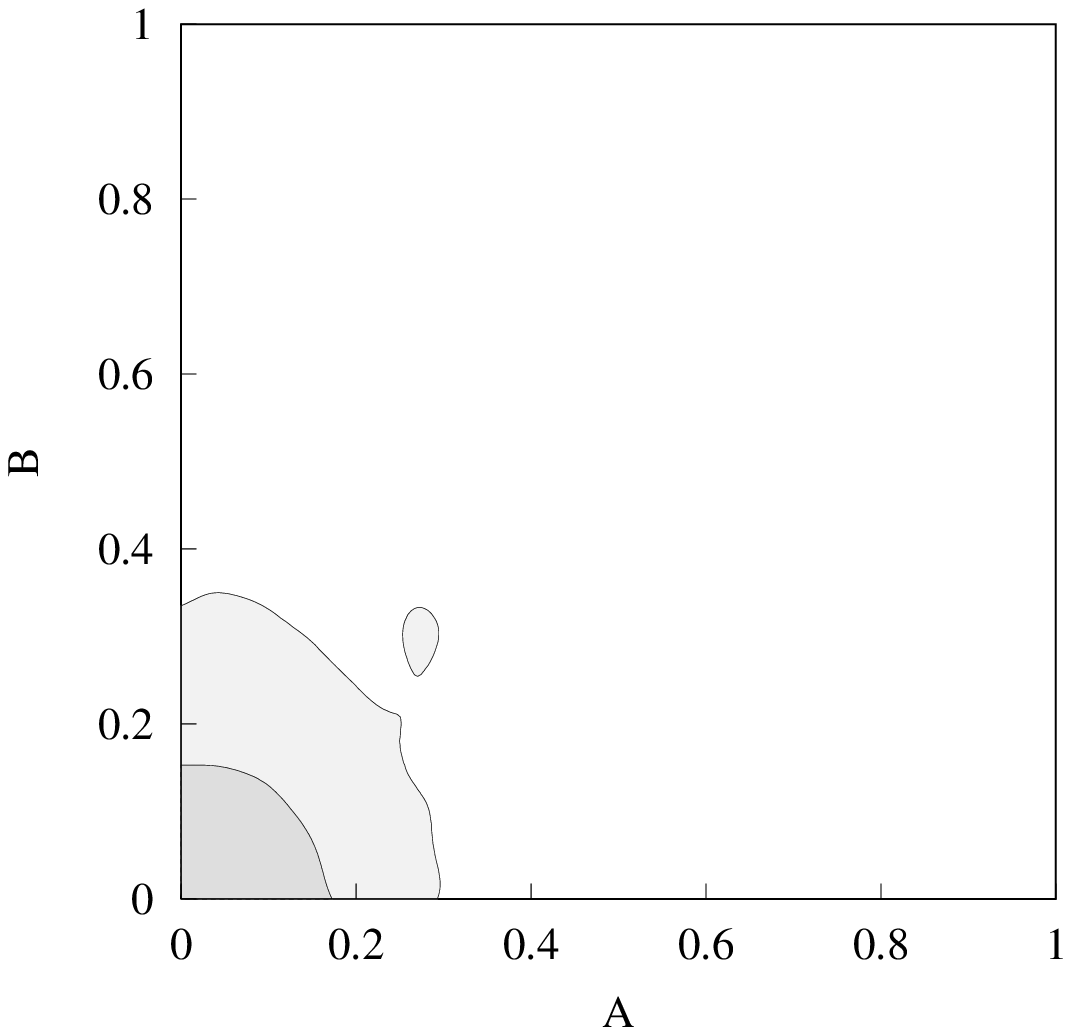}\includegraphics[width=6.8cm]{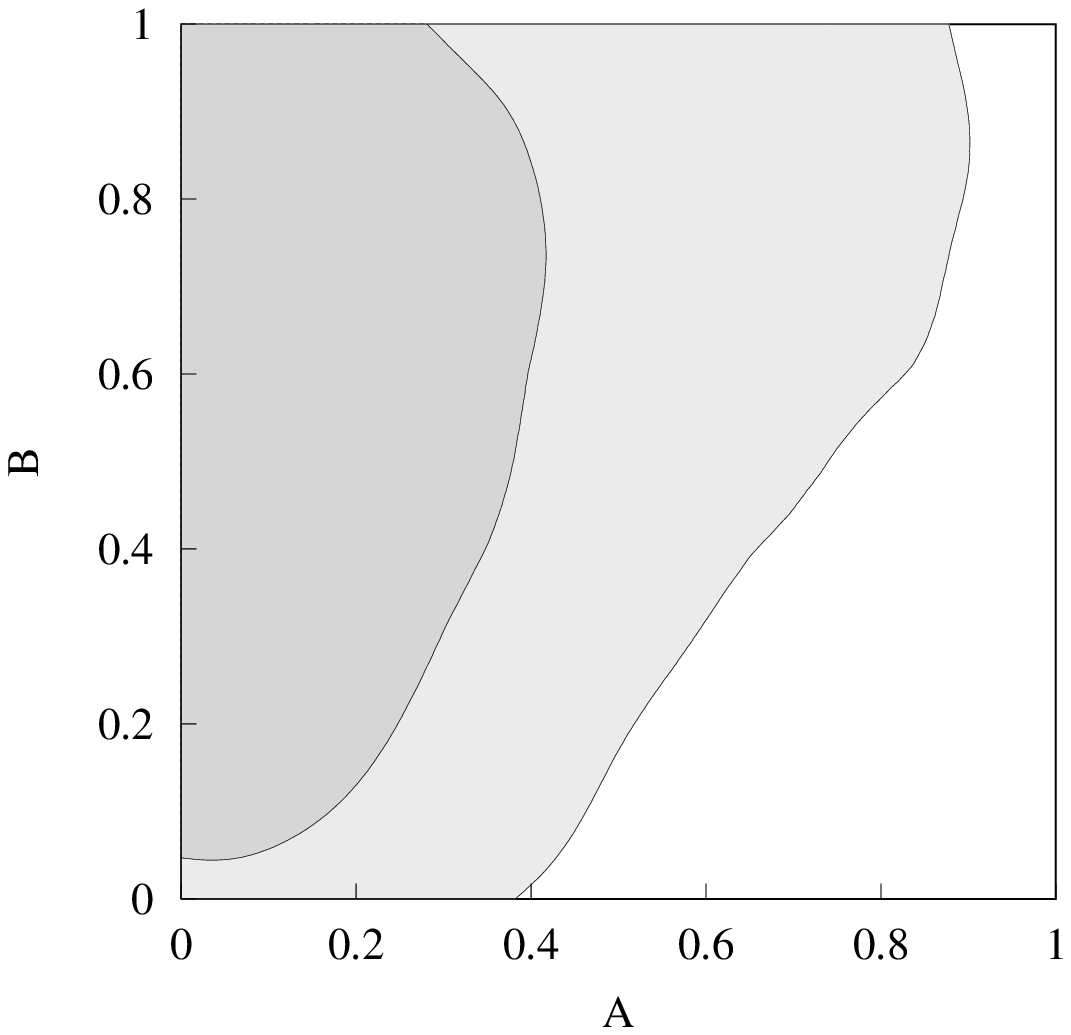}\\
\includegraphics[width=6.8cm]{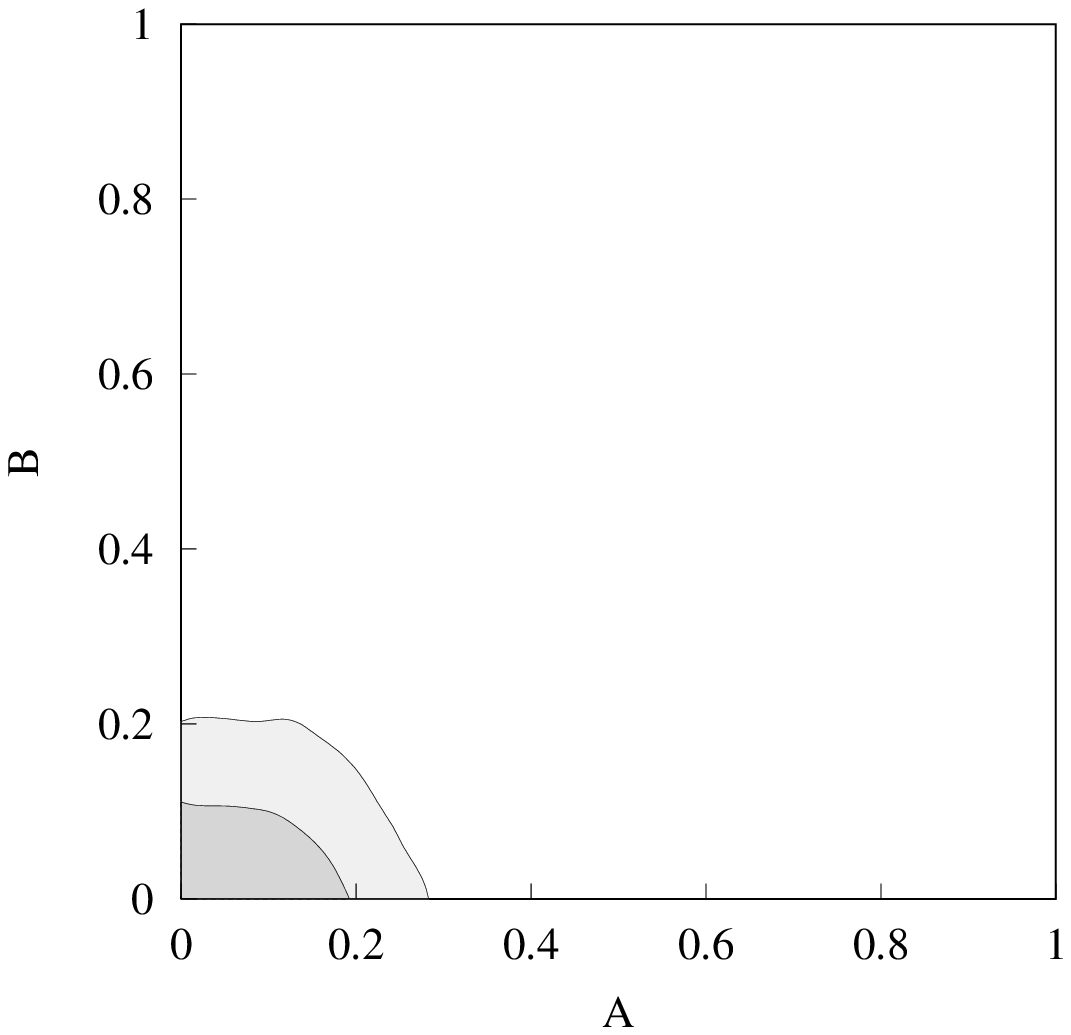}\includegraphics[width=6.8cm]{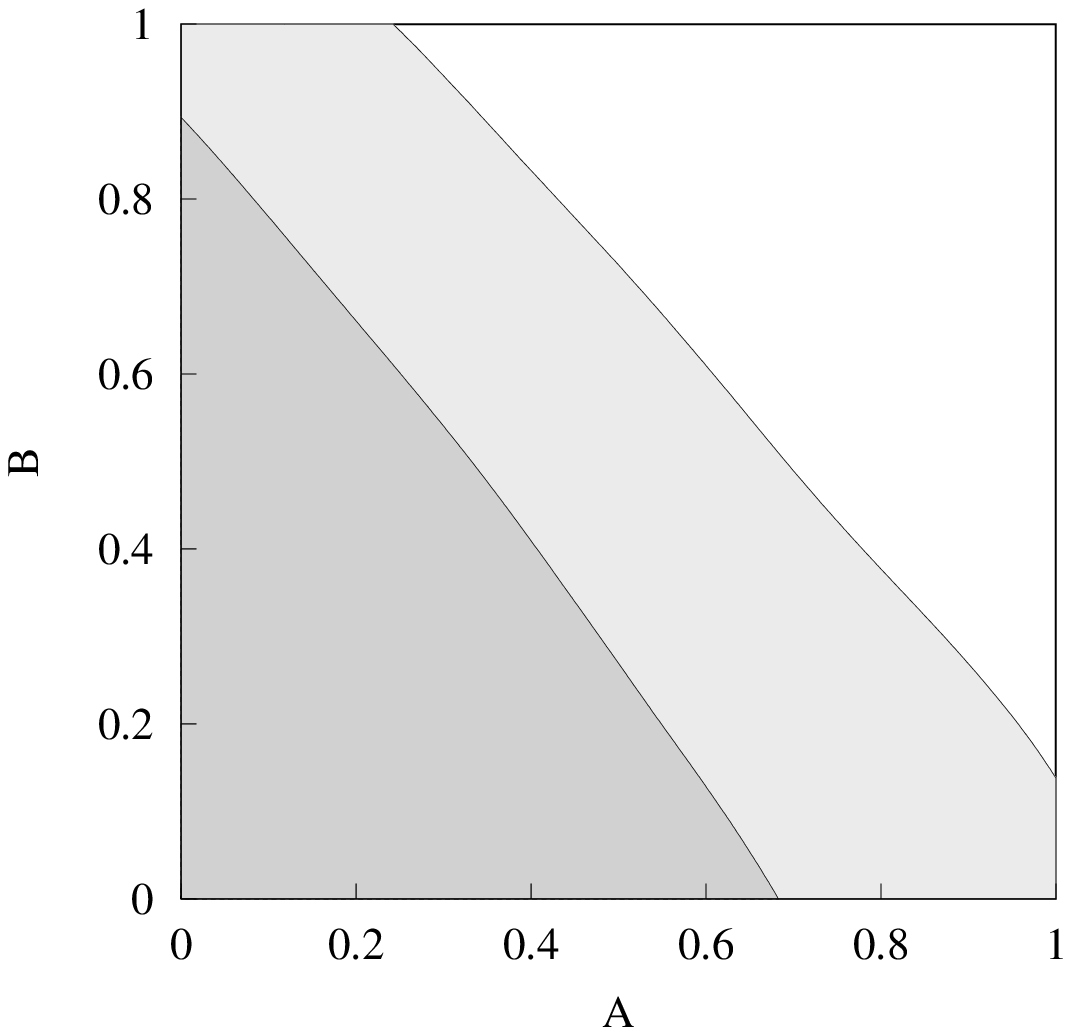}\\
\includegraphics[width=6.8cm]{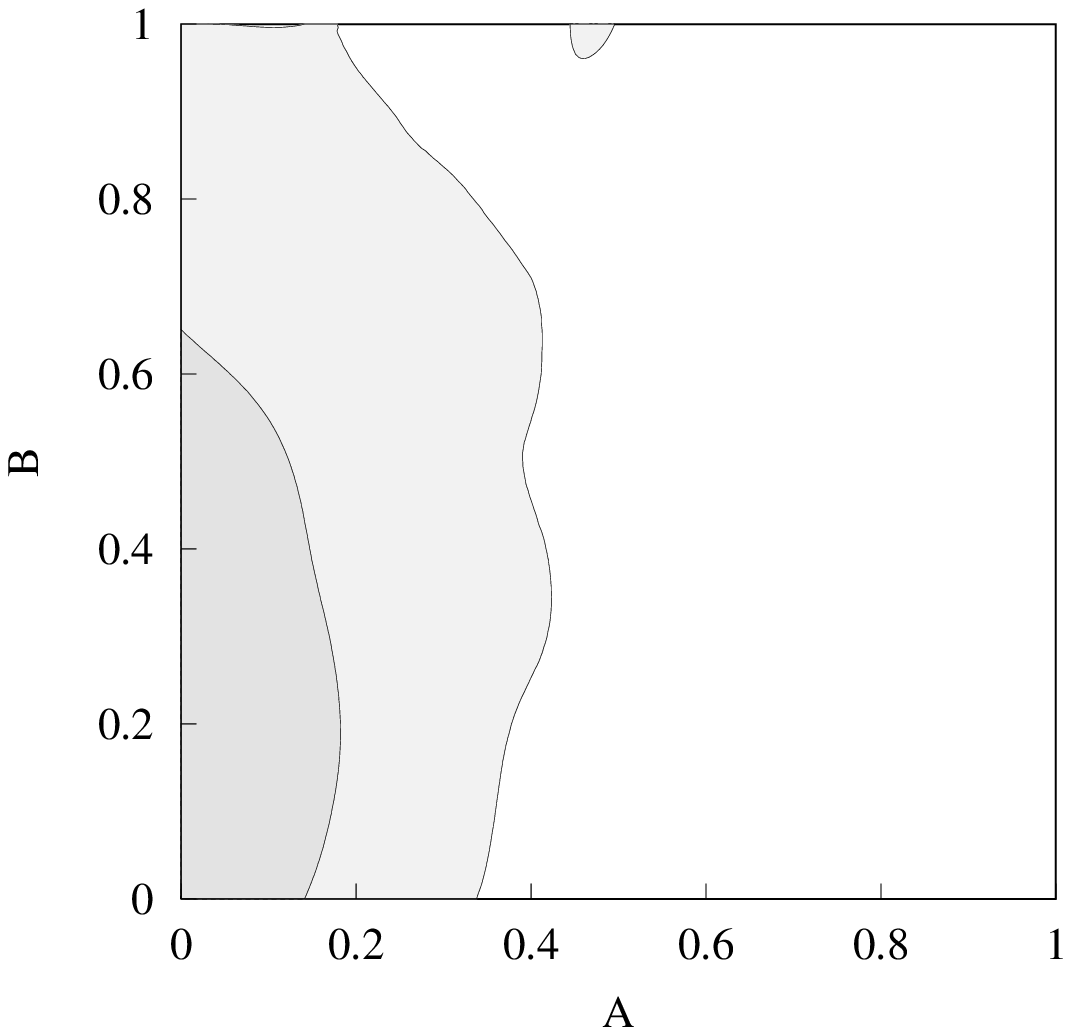}\includegraphics[width=6.8cm]{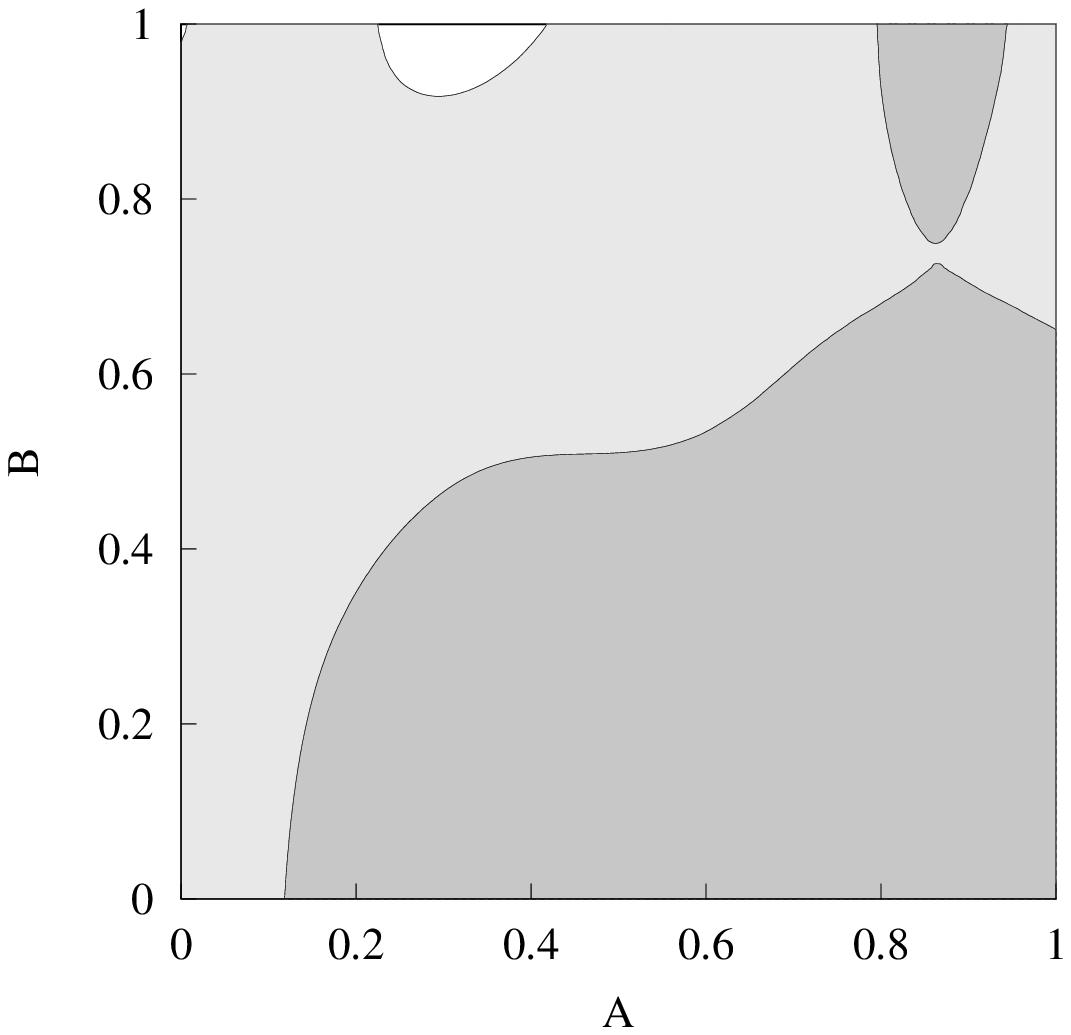}
\caption{\label{fig10} Top to bottom: marginalized likelihood in the $(A,B)$ plane for $N=2,3,4$ and $\a=0.5$ (left column) or $\a=0.1$ (right column).}
\end{figure}

Nevertheless, CMB observations do provide an interesting constraint on the highest value of $\a$ and, with some extra marginalization of the parameter space, also on the amplitudes $A$ and $B$ of the measure. Cases {\bf (vi)}--{\bf (xi)} are presented simultaneously in figures \ref{fig9} and \ref{fig10}.
When $\a$ is very small, the bounds on $A$ and $B$ are not particularly compelling for any $N$. On the other hand, for $\a=1/2$ the viable parameter space shrinks to a small portion (lower bounds $A,B>0$ are understood):
\be\label{boundsABstrong}
\boxd{\begin{matrix}
N=2:\qquad & A<0.3,\,B<0.4\,,\\
N=3:\qquad & A<0.3,\,B<0.2\,,\\
N=4:\qquad & A<0.4,\,B<1.0\,.
\end{matrix}}
\ee
Looking at figure \ref{fig9}, it is clear that the $\a=1/2$ case does not constrain $k_*$ well.
 

\section{Discussion}\label{disc}

In tables \ref{tab1}--\ref{tab4} we summarize the particle-physics, astrophysics and cosmological constraints obtained in \cite{frc8,frc12,frc13,qGW} and compare them with those of this paper. These tables represent the status of the phenomenology of multi-scale theories to date. Bounds obtained directly from experiments assuming $t_*$, $\ell_*$ and $E_*$ to be independent are in boldface. Bounds in normal font are obtained from the direct bounds by a unit conversion with Planck scales \cite{frc12,frc13}; for instance, knowing $E_*$ one can find $t_*=\tp\mpl/E_*$ and $\ell_*=\lp\mpl/E_*$. All figures are rounded and ``$\sim$'' indicates crude estimates. Items ``---'' are cases where the theory gives the standard result or where the experiments listed in the tables are unable to place significant constraints. Items ``?'' are cases not explored yet.


\subsection{Theory with weighted derivatives}

\begin{table}[ht]
\begin{center}
\begin{tabular}{|l|c|c|c|c|c|c|}\hline
Theory with weighted derivatives                  & $t_*$ (s)        & $\ell_*$ (m) & $E_*$ (eV)      & $A,B$ & source      \\\hline\hline
muon lifetime     																& --- 						 & --- 					& --- 						& --- 	& \cite{frc13}\\
Lamb shift       																  & ${\bf <10^{-23}}$& $<10^{-14}$  & $>10^7$    	    & --- 	& \cite{frc13}\\
measurements of $\a_{\rm QED}$ 										& ${\bf <10^{-26}}$& $<10^{-17}$  & $>10^{10}$      & ---		& \cite{frc13}\\\hline
$\frac{\Delta\a_{\rm QED}}{\a_{\rm QED}}$ quasars	& ${\bf <10^{11}}$ & $<10^{20}$   & $>10^{-28}$     & ---   & this paper\\
Gravitational waves 															& --- 						 & ---  				& --- 						& --- 	& \cite{qGW}\\
GRBs $\sim$ 																		  & --- 						 & ---  				& --- 						& --- 	& \cite{qGW}\\\hline
CMB black-body spectrum 													& ? 							 & ? 					  & ? 							& --- 	&   \\
CMB primordial spectra 													  & ? 							 & ? 					  & ? 							& ?	  	&  \\\hline
\end{tabular}
\caption{Absolute bounds on the hierarchy of multi-fractional spacetimes with weighted derivatives (obtained for $\a_0,\a\ll 1$). Bounds obtained directly from experiments assuming $t_*$, $\ell_*$ and $E_*$ to be independent are in boldface. Bounds in normal font are obtained from the direct bounds by a unit conversion with Planck scales. All figures are rounded and ``$\sim$'' indicates crude estimates. Items ``---'' are cases where the theory gives the standard result or where the experiments listed in the tables are unable to place significant constraints. Items ``?'' are cases not explored yet. The bound from quasars (fifth line) can be found from equations (1) and (73) of \cite{frc8}. Inverting (73), one finds $t_*\simeq t(|\Delta\a_{\rm QED}/\a_{\rm QED}|^{-1}-1)^{-1/(1-\a_0)}$, which has a maximum at $\a_0=0$. Like most upper bounds on multi-scale effects, including log oscillations would only weaken these bounds further.\label{tab1}}
\end{center}
\end{table}

\begin{table}[ht]
\begin{center}
\begin{tabular}{|l|c|c|c|c|c|c|}\hline
Theory with weighted derivatives                  & $t_*$ (s)     		& $\ell_*$ (m) & $E_*$ (eV)      & $A,B$ & source      \\\hline\hline
muon lifetime     																& --- 							& --- 				 & --- 						 & --- 	 & \cite{frc13}\\
Lamb shift       																  & ${\bf <10^{-29}}$ & $<10^{-20}$  & $>10^{13}$      & --- 	 & \cite{frc13}\\
measurements of $\a_{\rm QED}$ 										& ${\bf <10^{-36}}$ & $<10^{-27}$  & $>10^{20}$    	 & ---	 & \cite{frc13}\\\hline													
$\frac{\Delta\a_{\rm QED}}{\a_{\rm QED}}$ quasars	& ${\bf <10^6}$ 		& $<10^{15}$   & $>10^{-23}$     & ---   & \cite{frc8}\\
Gravitational waves 															& --- 							& ---  				 & --- 						 & --- 	 & \cite{qGW}\\
GRBs $\sim$ 																		  & --- 							& ---  				 & --- 						 & --- 	 & \cite{qGW}\\\hline
CMB black-body spectrum 													& $<10^{-21}$	  		& $<10^{-12}$  & ${\bf >10^2}$   & ---	 & this paper  \\
CMB primordial spectra 													  & ? 							  & ? 					 & ? 							 & ?	   & \\\hline
\end{tabular}
\caption{Bounds on the hierarchy of multi-fractional spacetimes with weighted derivatives for $\a=1/2=\a_0$.\label{tab2}}
\end{center}
\end{table}

The cosmological results for the theory with weighted derivatives are still rather limited. The $\a_0=1/2$ bounds from the CMB black-body spectrum are stronger than the very poor bound coming from the variation of the fine-structure constant in quasar signals \cite{frc8} but they are much weaker than any of the bounds found so far in particle physics. We could not find absolute bounds but we do not exclude them \emph{a priori}.

We have not derived the inflationary spectra of this theory either but we can make some preliminary remarks useful for future studies. The gravitational action of this theory takes the same form of the one in a Weyl integrable spacetime (WIST), the only difference being that the WIST dynamics features a scalar field that is replaced by a non-dynamical function of the measure here. After a conformal transformation of the metric, the WIST action is reduced to minimally coupled general relativity with an extra scalar field. In the theory with weighted derivatives, this extra part acts as a spacetime-dependent cosmological constant that does not fluctuate. To generate the inflationary spectra, one must include an inflaton field. The only difference with respect to ordinary cosmology would be in the effective mass of the Mukhanov--Sasaki equation for perturbations, which is determined by the background dynamics. The latter is given by the inflationary solutions of the Friedmann equations in the case of general relativity, while in the theory with weighted derivatives the Friedmann equations are replaced by two ``master equations'' describing the homogeneous and isotropic dynamics (see (5.43) and (5.44) of \cite{frc11}). The possibility to obtain acceleration directly from the anomalous geometry without imposing the slow-roll conditions might affect the inflationary spectra in some non-trivial way to be explored.


\subsection{Theory with \texorpdfstring{$q$}{}-derivatives}

\begin{table}[ht]
\begin{center}
\begin{tabular}{|l|c|c|c|c|c|c|}\hline
Theory with $q$-derivatives      			            & $t_*$ (s)         & $\ell_*$ (m) & $E_*$ (eV)      & $A,B$ & source      \\\hline\hline
muon lifetime    																  & ${\bf <10^{-13}}$ & $<10^{-5}$   & $> 10^{-3}$     & ---   & \cite{frc12}\\
Lamb shift        																& $<10^{-23}$       & $<10^{-15}$  & ${\bf >10^7}$   & ---   & \cite{frc12}\\
measurements of $\a_{\rm QED}$ 										& ---        				& ---          & --- 						 & ---   & \cite{frc13}\\\hline		
$\frac{\Delta\a_{\rm QED}}{\a_{\rm QED}}$ quasars	& ---		 					  & ---  				 & ---   					 & ---   & \cite{frc8}\\
Gravitational waves 															& $<10^{-22}$       & $<10^{-14}$  & ${\bf >10^{7}}$ & ---   & \cite{qGW}\\
GRBs $\sim$ 																			& $<10^{-32}$       & $<10^{-24}$  & ${\bf >10^{26}}$& ---   & \cite{qGW}\\\hline
CMB black-body spectrum 													& ? 							  & ? 					 & ? 							 & --- 	 &   \\
CMB primordial spectra 													  & $<10^{12}$      &${\bf <10^{20}}$& $>10^{-27}$	   & ---   & this paper  \\\hline
\end{tabular}
\caption{Absolute bounds on the hierarchy of multi-fractional spacetimes with $q$-derivatives (obtained for $\a_0,\a\ll 1$ in all cases but for the last one, where a likelihood analysis has been used).\label{tab3}}
\end{center}
\end{table}
\begin{table}[ht]
\begin{center}
\begin{tabular}{|l|c|c|c|c|c|c|}\hline
Theory with $q$-derivatives      			            & $t_*$ (s)     		& $\ell_*$ (m) & $E_*$ (eV)      & $A,B$ & source      \\\hline\hline
muon lifetime    																  & ${\bf <10^{-18}}$ & $<10^{-9}$   & $> 10^2$        & ---   & \cite{frc12}\\
Lamb shift        																& $<10^{-27}$       & $<10^{-19}$  & ${\bf >10^{11}}$& ---   & \cite{frc12}\\
measurements of $\a_{\rm QED}$ 										& ---        				& ---          & --- 						 & ---   & \cite{frc13}\\\hline											
$\frac{\Delta\a_{\rm QED}}{\a_{\rm QED}}$ quasars	& ---		 					  & ---  				 & ---   					 & ---   & \cite{frc8}\\
Gravitational waves 															& $<10^{-39}$       & $<10^{-30}$  & ${\bf >10^{23}}$& ---   & \cite{qGW}\\
GRBs $\sim$ 																			& $<10^{-50}$       & $<10^{-42}$  & ${\bf >10^{44}}$& ---   & \cite{qGW}\\\hline
CMB black-body spectrum 													& $<10^{-26}$       & $<10^{-18}$  & ${\bf >10^{10}}$& ---   & this paper \\	
CMB primordial spectra 													  & ---               & ---				   & ---		  			 & ${\bf <0.4}$ & this paper  \\\hline
\end{tabular}
\caption{Bounds on the hierarchy of multi-fractional spacetimes with $q$-derivatives for $\a=1/2=\a_0$.\label{tab4}}
\end{center}
\end{table}

In the case with $q$-derivatives, the bounds from the CMB black-body spectrum are much better than those from the muon decay time and are only one order of magnitude weaker than those from the Lamb shift. This marks a huge difference with respect to no-scale toy models, where particle-physics and atomic constraints are much stronger than cosmological ones \cite{CO}. In contrast with the theory with weighted derivatives, measurements of the fine-structure constant do not constrain the $q$-theory. Consequently, CMB black-body constraints are as competitive as those from particle physics.

Due to the strong degeneracy in the parameter space, the CMB inflationary spectra do not constrain the hierarchy of scales of the multi-fractal geometry in a significant way and, in fact, they fare much worse than the black-body spectrum. Making the identifications
\be\label{kell}
k_*=\frac{1}{\ell_*}\,,\qquad t_*=\frac{\tp}{\lp k_*}\,,\qquad E_*=\mpl\lp k_*\,,
\ee
from \Eq{lowbaa2} ($1\,{\rm Mpc}\approx 3\times 10^{22}\,{\rm m}$) we get the $\a$-independent bounds
\be
\ell_*<10^{20}\,{\rm m}\,,\qquad t_*< 10^{12}\,{\rm s}\,,
\ee
which are as weak as the bounds from $\Delta\a_{\rm QED}/\a_{\rm QED}$ in the theory with weighted derivatives. Using the Lamb-shift constraint $\ell_*<10^{-15}\,{\rm m}\sim 10^{-38}\,{\rm Mpc}$ in table \ref{tab3}, we get the rough lower limit
\be\label{k*min}
k_*> 10^{38}\,{\rm Mpc}^{-1}\,,
\ee
which is 36 orders of magnitude larger than \Eq{lowbaa2}. The bounds obtained from the gravitational waves observed in the black-hole merger GW150914 \cite{Abb16} for the theory with $q$-derivatives are much stronger than those from particle physics because they rely on a restriction of violation of Lorentz invariance in the propagation of gravitons \cite{qGW}. In this case, the lower bound \Eq{k*min} is further increased by 11 orders of magnitude, $k_*> 10^{52}\,{\rm Mpc}^{-1}$. The estimate from GRBs is based on a heuristic calculation and should not be taken \emph{verbatim} but it is expected to be roughly 10 orders of magnitude stronger than that from gravitational waves and 20 orders stronger than the Lamb-shift case. Thus, neither CMB nor particle physics can compete with Lorentz-violation constraints in general. The bound on $k_*$ from the CMB black-body spectrum is only one order of magnitude weaker than \Eq{k*min}. If we had taken this result from \cite{frc12} on board at the outset, we could have concluded in a few lines that the CMB scales are insensitive to the typical hierarchy of multi-fractional geometries. 

However, both the gravitational-wave and the GRB bounds have a scope more limited than the CMB analysis performed here and they should be regarded as complementary to (rather than superseding) the other constraints. On one hand, they do not apply to the theory with weighted derivatives \cite{qGW}. On the other hand, in the case with $q$-derivatives they can be relaxed considerably either for $\a\neq 1/2$ (as done in table \ref{tab3}) or in the presence of log oscillations and they can be evaded. In contrast, small changes in $\a$ do not influence much the CMB analysis presented here. To confront a theory against experiments, it is fundamental to obtain independent constraints coming from as many sectors of physics a possible. Such was one of the goals of this paper.

Fortunately, and perhaps contrary to expectations given the weakness of the above results, information from the CMB primordial spectra does not stop on the scale hierarchy and, in fact, it provides a unique insight into those directions in the parameter space not covered by previous analyses: the average fractional exponent $\a$ in the spatial direction (unconstrained in \cite{frc8,frc12,frc13}) and the parameters of the log-oscillatory part of the measure, which describes the fundamental discrete structure of multi-scale spacetime. Let us discuss these results.

To begin with, we have obtained the lower bound \Eq{lowbaa} on $\a$, which should be interpreted with care. Provided \Eq{kell} is true (which we can always assume to be the case, in order to simplify the theory), the constraint \Eq{lowbaa} has been obtained in a region in the parameter space already excluded by observations (compare \Eq{k*min} with figure \ref{fig6}). In other words, we cannot use the inequality \Eq{lowbaa} as an absolute lower bound on $\a$.

However, we also found a robust upper bound on $\a$ in the presence of log oscillations, given by \Eq{lowbaa2}. Recalling that the Hausdorff dimension of these spaces is $\dh^{\rm \,space}\simeq (D-1)\a$ in the UV \cite{frc1,frc7}, for $D=4$ topological dimensions the first of the two constraints in \Eq{lowbaa2} translates into the upper bound \Eq{dhst} on the UV spatial dimension, which is as low as $\dh^{\rm \,space}<0.3$ for $N=2$. Since we have not constrained the fractional exponent $\a_0$ in the time direction, we cannot say anything about the dimension of spacetime. Inequality \Eq{lowbaa2} is compatible with the range of $\a$ for which GRB hierarchy constraints place the characteristic energy scale of the theory below the Planck scale. Having $E_*<\mpl$ is a necessary condition to have a self-consistent hierarchy where the fundamental scale in log oscillations is at the top. Therefore, we have gathered more evidence that the theory is viable only for small $\a$.

Finally, fixing $\a$ we were able to constrain the size of the amplitudes in the oscillatory part of the multi-fractional measure (figure \ref{fig10}). These results are presented in \Eq{boundsABstrong}. The larger $\a$, the more stringent are the upper bounds on $A$ and $B$ and the case $\a=1/2$ is already tightly limited.

Having completed the survey of the parameter space of the theories with weighted and $q$-derivatives with a wealth of heterogeneous experiments, it is time to take stock of what we know and plan for future work. There are some small gaps to fill in tables \ref{tab1}--\ref{tab2}: we do not have absolute bounds from the CMB black-body spectrum, nor bounds from primordial spectra in the case of weighted derivatives. However, it is likely that these data will not add much to the discussion. For instance, absolute bounds are usually weaker than $\a=1/2$ bounds. The CMB primordial spectra in the theory with weighted derivatives might be more interesting and reserve some surprises about the viable ranges of $\a$, $A$ and $B$. Still, it is becoming clear that the phenomenology of both multi-fractional theories is reaching its limit. On the other hand, we know nothing about the phenomenology of another multi-scale theory, the one with fractional derivatives. This theory is more promising than the two studied here because, contrary to the latter \cite{frc9}, it might be able to quantize gravity perturbatively \cite{frc2}. However, its stage of formal development is still preliminary and it will take some time to extract some phenomenology. We hope to do so in the near future.


\begin{acknowledgments}
The work of G.C.\ is under a Ram\'on y Cajal contract and is supported by the I+D grant FIS2014-54800-C2-2-P. S.K.\ is supported by the Career Development Project for Researchers of Allied Universities. S.T.\ is supported by the Grant-in-Aid for Scientific Research Fund of the JSPS Nos.\,24540286, 16K05359, the MEXT KAKENHI Grant-in-Aid for Scientific Research on Innovative Areas ``Cosmic Acceleration'' (No.\,15H05890) and the cooperation program between Tokyo University of Science and CSIC.
\end{acknowledgments}


\end{document}